\documentclass[final,3p,times,authoryear]{elsarticle}

\usepackage{amssymb}
\usepackage{lipsum}
\usepackage{amsmath}
\usepackage{amssymb}
\usepackage{textcomp}
\usepackage{graphicx}
\usepackage{caption}
\usepackage{subcaption}
\usepackage{array}
\usepackage{textcomp}
\usepackage{mathtools,cuted}
\usepackage{centernot}
\usepackage{slashed}
\usepackage{hyperref}
\usepackage{xcolor}
\usepackage{natbib}

\journal{Nuclear Physics B}

\begin{document}

\begin{frontmatter}

\title{Interplay of type-I and type-II seesaw in neutrinoless double beta decay in left-right symmetric model}

\author[mymainaddress]{Vivek Banerjee}
\ead{vivek$\_$banerjee@nitrkl.ac.in}

\author[mymainaddress]{Sasmita Mishra}
\ead{mishras@nitrkl.ac.in}

\address[mymainaddress]{Department of Physics and Astronomy, National Institute of Technology Rourkela, Sundargarh, Odisha, India, 769008}

\begin{abstract}
	The left-right symmetric models (LRSM) generally include type-I and type-II induced seesaw masses
	as a hybrid mass for the light-active neutrinos. Assuming a particular form of Dirac-type coupling,
	the Majorana-type coupling present in the seesaw mass formula can be expressed in terms of
	low-energy neutrino oscillation observables and vacuum expectation values (vevs) of the scalar fields
	present in the model. The Majorana-type coupling thus admits eight different solutions by
	considering whether the type-I and type-II terms dominate the light neutrino mass.
	We study the role of all eight solutions in the lepton number violating neutrinoless double beta decay
	($0\nu\beta\beta$) process. In LRSM, the right-handed neutrinos, triplet scalars, and gauge bosons
	of the left and right sectors act as mediators of new contributions to the $0\nu\beta\beta$ process.
	As a result, the effective mass of electron neutrino appearing in the decay width would be a
	function of $v_R$ (vev of the Higgs triplet of the right sector) along with other
	parameters of the model, through the masses of the new contributions. The energy scale,
	$v_R$ can be considered as the new physics scale which allows exploring physics beyond the Standard Model.
	Considering the present and future sensitivity of searches of $0\nu\beta\beta$, we study the
	role of eight different solutions of the Majorana coupling matrix. In our study, the inverted
	hierarchy of light neutrino masses is disfavored for all solutions keeping future sensitivity of
	effective mass in the picture, if the lightest mass of active neutrinos is below $0.001$ eV.
	Also, our study shows a possibility of new physics contributions saturating
	the experimental bound on effective mass for $v_R$ in the range of $10$ TeV for two particular solutions
	of the Majorana coupling matrix and simultaneously provides the insights about parity breaking scale.
\end{abstract}
\end{frontmatter}

\section{Introduction}
\label{sec:intro}
The famous framework of neutrino mass generation via 
seesaw mechanism \cite{Giunti:2007ry} is the Standard Model (SM) extended with 
heavy singlet right-handed
neutrinos. One needs more than just one right-handed neutrino to 
account for the light neutrino mass generation, compatible with experimental 
data. Apart from right-handed neutrinos, scalar and fermionic triplets 
is also potential, beyond SM, candidates that can generate
light neutrino mass via seesaw mechanism. Nevertheless, working in a model
which is theoretically motivated has its advantages. For example, the addition of
right-handed neutrinos and scalar triplets are also consistent
with the models inspired by grand unification such as 
left-right symmetry \cite{Mohapatra:1974gc},\cite{Senjanovic:1975rk},\cite{Senjanovic:1978ev},
and $SO(10)$ \cite{Georgi:1974my,Fritzsch:1974nn}. 
In such models, heavy fields such as scalar triplets and 
fermion triplets (singlets) arise, based on the symmetry arguments, and can account for
light neutrino mass generation via seesaw mechanism. In LRSM, after parity breaking and subsequent electroweak symmetry breaking, the light neutrino mass gets
generated from combination of type- I \cite{Minkowski:1977sc,Ramond:1979py,Gell-Mann:1979vob,Sawada:1979dis,Levy:1980ws,Mohapatra:1979ia} and type -II \cite{Magg:1980ut,Lazarides:1980nt,	Mohapatra:1980yp,Schechter:1980gr} seesaw mechanisms.
The experimental
verification of the underlying theory at the seesaw scales is an important
question which is worth exploring. 
In this work, we study the effect of the interplay of type-I and type-II seesaw
mass contributions to the light neutrino mass in the LRSM, in exploring the seesaw scale using low energy 
phenomenology of  $0\nu\beta\beta$ decay.

Originally
proposed to explain the parity violation in the SM, the LRSM
are predictive in many ways. 
In the SM the knowledge of charged fermion masses can uniquely determine
the corresponding Dirac Yukawa coupling implying the branching ratio
$(\Gamma(h\rightarrow f\bar{f})\propto m_f^2$). However, making similar
predictions for neutrinos is still not possible in the domain of the SM.
In type -I seesaw framework, assuming the heavy and light neutrino masses and light neutrino flavor mixing measured uniquely, the Dirac Yukawa coupling can be determined 
using Casas-Ibarra \cite{Casas:2001sr} parametrization. But, the presence of an arbitrary orthogonal matrix introduces some ambiguity and the knowledge of the origin of neutrino mass remains unknown. However, in the LRSM, the Dirac Yukawa coupling can in principle be  determined as proposed by the authors of references \cite{Nemevsek:2012iq} and \cite{Senjanovic:2018xtu}. A recent analysis can be found in Ref.\cite{Kiers:2022cyc}. Similar
to Casas-Ibarra parametrization, in this case, the authors assume along with
light neutrino sector the right-handed heavy neutrino masses and the corresponding
mixing can be probed in the experiments \cite{Keung:1983uu}.  As a result, the Dirac Yukawa coupling of neutrinos can be uniquely determined and one can make predictions for both high and low energy phenomena such as decays of heavy neutrino, electric dipole moments of charged leptons, neutrinoless double beta decay, etc. Although in LRSM the light neutrino mass is of both type-I and type-II origin, the preceding analyses only take type-I contribution into account. Unlike Dirac type coupling,  Majorana type coupling can not be constrained experimentally. Nevertheless, the seesaw formula including
both type-I and type-II contributions can be employed to reconstruct the Majorana
coupling matrix, taking certain quantities as input parameters \cite{Akhmedov:2006de}. The study can provide some insight into the underlying theory at the 
seesaw scale.

Similar to the electroweak symmetry breaking
via Higgs mechanism in the SM, in LRSM the breaking of parity is achieved
by the  right-handed triplet scalar acquiring vacuum expectation value (vev). The
vev, $v_R$, of the triplet scalar is the
new physics scale. LRSM continues to be one of the simplistic extensions of the SM and 
experiments like LHC are looking for a trace of a new gauge boson. But the ATLAS detector 
excludes a $SU(2)_R$ gauge boson, $W_R$ mass smaller than $3.8 - 5$ TeV
in right-handed neutrino mass range $0.1 - 1.8$ TeV \cite{ATLAS:2019isd}. Hence, it is important to 
explore the scale of left-right symmetry breaking at energy ranges 
beyond TeV scale. 

In LRSM, the type-I and type-II contributions to light neutrino mass are
inherently related as both contributions depend on the same
Majorana type coupling matrix.
In phenomenological studies based on LRSM, one often assumes the dominance of
either type-I or type-II seesaw contribution to light  neutrino mass, while assuming specific values of unknown seesaw parameters. But considering both contributions comparable, 
the interplay of type-I and type-II seesaw can have interesting implications
for the interpretation of neutrino data. One attempt in this direction has been taken in Ref.\cite{Akhmedov:2006de} where
considering both contributions comparable, the authors show that the seesaw equation in the
LRSM has eight solutions for the eigenvalues of the  Majorana coupling matrix, $f$.
In LRSM, the eigenvalues of the  mass matrix of the right-handed neutrinos, $M_R = v_R f$, hence allow eight solutions.
By constructing the allowed structures of $M_R$ analytically, using low energy
neutrino mixing data, the dependence of $M_R$ on the mass spectrum of light
neutrinos, mixing angle $\theta_{13}$, leptonic CP violation, scale of 
left-right symmetry breaking and the hierarchy in the neutrino Yukawa 
couplings were studied in Ref. \cite{Akhmedov:2006de}. In a different approach, taking the Dirac mass matrix as that for up-quarks, a reconstruction procedure for the right-handed neutrino  masses has been conducted in Ref.\cite{Hosteins:2006ja}. In a similar set-up for Dirac mass matrix for neutrinos in type-I seesaw framework, and by introducing its dependence through renormalization group evolution, the authors of Ref.\cite{Pascoli:2003uh} show a common origin high energy CP violation and low energy effects such as lepton flavor violation, effective mass of electron neutrino in $0\nu\beta\beta$  decay processes and CP violation in neutrino oscillation. Also, in a similar set-up for Dirac type mass matrix for neutrinos, in a model based on $SO(10)$, the authors of Ref.\cite{Buccella:2017jkx} show its impact on the effective mass parameter of $0\nu\beta\beta$ decay process. The latter shows that  the normal hierarchy
is only allowed and the lower limit on the lightest neutrino mass, 
$m_{\rm lightest}\gtrsim 7.5 \times 10^{-4}$ eV at $3~\sigma$ level in compliance with predictions of effective mass parameter in standard three neutrino picture. 
In Ref.\cite{Hosteins:2006ja} the authors have developed a method for reconstructing the Majorana coupling matrix $f$ using the seesaw formula in the left-right symmetric model.
The $8$ solutions of $f$ can be employed to study phenomenology where the presence of right-handed 
neutrinos and triplet scalars play a role. In Ref.\cite{Hosteins:2006ja} using the $8$ solutions 
the authors have studied the implications of the  different set of 
solutions in lepton flavor violation  and flavor effects in leptogenesis from the decay of right-handed neutrinos. In this work we intend to study the phenomenology of neutrinoless double beta decay 
in left-right symmetric model where the  right-handed neutrinos and scalar triplets are potential candidates of new physics contribution. As pointed out in Refs.\cite{Hirsch:1997tr,Dvali:2023snt}
the process is a promising probe of new physics and can be entangled with collider searches, in particular in the search of new gauge bosons of left-right symmetry.
In this work, by taking three-generation case the Majorana
coupling matrix into consideration, we study the effects of the eight solutions, stated above, in
$0 \nu\beta \beta$ decay taking $v_R$ as the new physics scale.  

The paper is organized as follows. In section (\ref{sec:lrsm}), we summarize the 
LRSM in the context of generation of light neutrino masses via type-I and type-II
seesaw mechanism. In section (\ref{sec:dom_types}), we discuss the eight solutions 
of Majorana coupling matrix and the corresponding right-handed neutrino mass, $M_R$ 
as a function of Dirac type Yukawa couplings, lightest neutrino mass, and triplet scalar vev, $v_R$. 
The new contributions to $0\nu\beta\beta$ decay process in LRSM along with the light neutrino
contribution are discussed in section (\ref{sec:onbb}). In section (\ref{sec:calc}),
we discuss the numerical results. The conclusions of the study are summarized in section (\ref{sec:concl}).
\section{Left-right symmetric model}
\label{sec:lrsm}
The gauge group of the left-right symmetric models is a simple extension of the
SM gauge group given by $SU(3)_c\times SU(2)_L\times SU(2)_R\times U(1)_{B-L}$. Originally
proposed to explain spontaneous breaking of parity, it requires right-handed (RH) neutrino
degrees of freedom.  The right-handed neutrinos give rise to Majorana masses and naturally 
explain the light neutrino masses.  In the minimal model, the  doublets of quarks and leptons are represented as, with their gauge charges,
\begin{equation}
Q_{L,i}: \left(3,2,1,\frac{1}{3}\right) \equiv 
\begin{pmatrix} 
u_L \\ d_L  
\end{pmatrix}_i, 
\quad Q_{R,i}: \left(3,1,2, \frac{1}{3}\right) \equiv
\begin{pmatrix} 
u_R \\ d_R  
\end{pmatrix}_i , 
\end{equation}
\begin{equation}
l_{L,i} \left(1,2,1, -1\right) \equiv
\begin{pmatrix} 
\nu_L \\ e_L  
\end{pmatrix}_i ,
\quad 
l_{R,i}: \left(1,1,2, -1\right)\equiv
\begin{pmatrix} 
\nu_R \\ e_R 
\end{pmatrix}_i ,
\end{equation}
where $i =1,2,3$ represents the family index. The subscripts $L$ and $R$ denote the left- and right-handed chirality, with the chiral projection
operators, $P_{L,R} = (1\mp \gamma_5)/2$ respectively. The $B-L$ charges are
assigned using the formula for electric charge,
\begin{equation}
Q = I_{3L} + I_{3R} +\frac{B-L}{2}.
\end{equation}
The Higgs content \cite{Maiezza:2016ybz} of this model is much richer than the SM. There is a Higgs bidoublet $\phi$ to give masses to all fermions through Yukawa type interaction. The 
gauge transformation properties of the fermion fields require $\phi$ and $\tilde{\phi} \equiv
\tau_2 \phi^* \tau_2$ transform as $(1,2,2,0)$. It is represented as \cite{Abada:2007ux},
\begin{equation}
\phi = 
\begin{pmatrix} 
\phi_1^0 & \phi_1^+ \\ \phi_2^- & \phi_2^0 
\end{pmatrix}.
\end{equation}
In order to break the LRSM gauge group to the SM gauge group, the Higgs sector has to be enlarged. Although there is not an unique way to achieve this objective, the interesting models are obtained by introducing scalar triplets represented as,
\begin{equation}
\Delta_{L}: (1,3,1,2) \equiv
\begin{pmatrix} \frac{\delta_{L}^+}{\sqrt{2}} 
& \delta_{L}^{++} \\ \delta_{L}^0 & \frac{\delta_{L}^+}{\sqrt{2}}  
\end{pmatrix}, \quad
\Delta_{R}: (1,1,3,2) \equiv
\begin{pmatrix} \frac{\delta_{R}^+}{\sqrt{2}} 
& \delta_{R}^{++} \\ \delta_{R}^0 & \frac{\delta_{R}^+}{\sqrt{2}}  
\end{pmatrix}.
\end{equation}
The scalar triplets have Yukawa type couplings with the lepton with lepton number
$L =-2$.
The left-right symmetric models with not more than three scalar multiplets 
$\phi, \Delta_L$, and $\Delta_R$ are called minimal left-right symmetric model (MLRSM).
Under the weak and $B-L$ sectors the fermions $\psi_{L(R)}$ and the scalar multiplets transform as shown in table.(\ref{tab:group}),
\begin{table}[h]
	\centering
	\setlength{\tabcolsep}{10pt} 
	\renewcommand{\arraystretch}{1.2}
	\begin{center}
		\begin{tabular}{||c | c||} 
			\hline
			$SU(2)_L\times SU(2)_R$ & 	$U(1)_{B-L}$ \\ [1.0ex] 
			\hline\hline
			$\psi_{L(R)} \rightarrow U_{L(R)} \psi_{L(R)}$ & $\psi_{L(R)} \rightarrow e^{i \theta_{B-L}} \psi_{L(R)}$  \\ 
			\hline
			$\phi \rightarrow U_{L} \phi U^\dagger_R$ & $\phi \rightarrow \phi$  \\ 
			\hline
			$\Delta_{L(R)} \rightarrow U_{L(R)} \Delta_{L(R)} U^\dagger_{L(R)}$ & $\Delta_{L(R)} \rightarrow e^{i \theta_{B-L}} \Delta_{L(R)}$  \\ 
			\hline
		\end{tabular}
	\end{center}
	\caption{Transformation of fermion and scalar fields under the gauge groups of LRSM. }
	\label{tab:group}
\end{table}

The Yukawa part of the total Lagrangian for leptons is given by,
\begin{equation}
L_Y = h_{ij}\bar{l^i}_L\phi l_R^j + g_{ij}\bar{l^i}_L\tilde{\phi}l_R^j +  i(f_L)_{ij}l_L^{iT}C\tau_2\Delta_L l_L^j + i(f_R)_{ij} l_R^{iT}C\tau_2\Delta_R l_R^j + h.c .
\label{eq:yukawa}
\end{equation}
Here $h$ and $g$ are the Dirac type Yukawa coupling constants. And,
$f_L$, $f_R$ are  Majorana type Yukawa couplings of left and right sectors respectively. 

In addition, the model has a discrete left-right symmetry, which can be implemented either
by parity, $P$ or the charge conjugation, $C$. Here we consider $P$ as the operator under
which the fields transform as,
\begin{equation}
\psi_L\longleftrightarrow \psi_R,\quad \Delta_L\longleftrightarrow\Delta_R,
\quad \phi\longleftrightarrow \phi^\dagger.  
\end{equation}
Requiring the Lagrangian is symmetric under parity as the left-right symmetry, 
the Yukawa couplings are constrained as, $(h_{ij} = h^*_{ji})$ and $(g_{ij} = g^*_{ji})$, and $f_L$ = $f_R$. Henceforth, we use $f_L =f_R =f.$

The gauge symmetry of the MLRSM breaks into the SM gauge group by the scalar fields with a desired vacuum alignment to have an electric charge conserving minima of the scalar potential \cite{Senjanovic:1978ev,Mohapatra:1980yp,Mohapatra:1979ia}, 
\begin{equation}
\langle \phi \rangle =
\begin{pmatrix}
\frac{k_1}{\sqrt{2}} & 0 \\ 
0 & \frac{k_2}{\sqrt{2}} 
\end{pmatrix}, \quad 
\langle \Delta_{L,R} \rangle =
\begin{pmatrix} 0 & 0 \\
\frac{v_{L,R}}{\sqrt{2}} & 0 
\end{pmatrix} .
\label{eq:vevs}
\end{equation}
Here the $k_{1,2}$ and $v_{L,R}$ are vacuum expectation values (vevs) of Higgs bidoublet and triplets respectively. Here we assume the order of magnitude relations,
\begin{equation}
|v_L|^2 \ll |k_1^2| + |k_2^2| \ll |v_R|^2.
\end{equation}
Taking only the effects of $v_R$ into account, it can be shown that 
$v_R$ breaks $SU(2)_R\times U(1)_{B-L}$ to $U(1)_Y$. The complete symmetry breaking pattern can be shown as,
\begin{equation}
SU(2)_L\times SU(2)_R\times U(1)_{B-L} \xrightarrow{v_{R}} SU(2)_L\times U(1)_Y
\xrightarrow{k_1, k_2} U(1)_{\rm em}.
\end{equation}
As far as gauge boson masses are concerned, there is an interesting relation between the mass of the gauge boson of left sector, $W_L$ and
vevs of the Higgs bidoublet, $M_{W_L}^2 =\frac{1}{4}g_L^2(k_1^2 + k_2^2)$. The mass of the gauge bosons of the right sector, after the parity symmetry breaking, is $M_{W_R}\simeq g_R v_R$.
Here $g_L$ and $g_R$ are the gauge coupling constants for left and right sector gauge bosons respectively. Following the group structure of MLRSM, the relation between gauge couplings is expressed as,
\begin{equation}
\frac{1}{e^2} = \frac{1}{g_L^2} + \frac{1}{g_R^2} + \frac{1}{g_{B-L}^2}.
\end{equation}  
\subsection{Neutrino mass}
\label{ssec:neu_mass}
Using the vevs given in Eq.(\ref{eq:vevs}), the mass matrix for the for charged leptons becomes
\begin{equation}
m_{l}=\frac{1}{\sqrt{2}}(hk_2 + gk_1).
\end{equation}
But for the neutrinos, the scenario is a bit different. The Dirac and Majorana both couplings are present in the Yukawa Lagrangian Eq.(\ref{eq:yukawa}). So, the mass term for neutrinos is a mixture of Dirac and Majorana type masses. Through a change of basis,
\begin{equation}
\nu=\frac{\nu_L + \nu_L^c}{\sqrt{2}}, \quad N=\frac{\nu_R + \nu_R^c}{\sqrt{2}},
\end{equation}
where the conjugate of a field is defined as $\psi^c = C\bar{\psi}^T$, and
after electroweak symmetry breaking the neutrino mass matrix in the basis of ($\nu,N$) becomes,
\begin{equation}
M_{\nu} = \begin{pmatrix}
fv_L & y_D v_H \\ y_D^T v_H & fv_R
\end{pmatrix},
\label{eq:mass}
\end{equation}
where the Dirac type Yukawa coupling for neutrino $y_D$ \cite{Deshpande:1990ip} is given as, 
\begin{equation}
y_D =\frac{1}{\sqrt{2}} \bigg(\frac{hk_1 +gk_2}{v_H}\bigg),
\end{equation}
where $v_H = \sqrt{k_1^2 + k_2^2} $ is the vev of the SM Higgs field.
Considering three generations of leptons, this mass matrix gets the dimension of $6\times6$, which can be represented as,
\begin{equation}
M_{\nu} = 
\begin{pmatrix}
m_L & m_D \\ m_D^T & m_R 
\end{pmatrix}_{6\times6}.
\end{equation}
The full neutrino mass matrix can be diagonalized through an unitary matrix U as
\begin{equation}
U^T M_{\nu} U = U^T 
\begin{pmatrix} m_L & m_D \\
m_D^T & m_R\end{pmatrix} U = 
\begin{pmatrix}
m_{\text{light}} & 0 \\ 0 & M_{\text{heavy}}
\end{pmatrix}.
\end{equation}
This unitary matrix U can be decomposed as $U=VU_\nu$   following the recursive decoupling method shown in \cite{Grimus:2000vj},
\begin{equation}
\begin{split}
U   & =
\begin{pmatrix} 
1 -\frac{1}{2}RR^\dagger  & R \\ 
-R & 1 -\frac{1}{2}R^\dagger R
\end{pmatrix} 
\begin{pmatrix} 
U_L &  0 \\
0 & U_R
\end{pmatrix} \\ 
 & =
 \begin{pmatrix} 
U_L -\frac{1}{2}RR^\dagger U_L & RU_R \\
-RU_L & U_R -\frac{1}{2}R^\dagger RU_R
\end{pmatrix} \\
& =
\begin{pmatrix} 
U_L' & T \\ S & U_R'  
\end{pmatrix}.
\end{split}
\end{equation}
The matrix $R$ is represented as, $R=m_D^{\dagger}M_R^{-1}$. It should be noted that
if the Dirac-type mass is absent in Eq. (\ref{eq:mass}), the light and heavy
neutrino sectors are completely decoupled. In this case $R=0$. So non-zero
value of $R$ signifies the light and heavy neutrino mixing.
$U_\nu$ is the block diagonal $6\times6$ mixing matrix, $U_{\nu} 
= \text{Diag}(U_L, U_R)$. Here $U_L$ and $U_R$ are the diagonalizing matrices of
light and heavy neutrinos respectively. Some of the works in literature have used $U_L$ = $U_R$ for phenomenological calculation \cite{Nemevsek:2012iq},\cite{Chakrabortty:2012mh}. 
The $U_R$ matrix is the diagonalizing matrix of the right-handed neutrinos which can be
obtained numerically, in our work, from the diagonalizing matrix of $f$ matrix (as 
$M_R = v_R f$).
In the present work, we have taken $U_L$ as the Pontecorvo-Maki-Nakagawa-Sakata (PMNS) matrix, $U_{\text{PMNS}}$ times a diagonal matrix involving Majorana phases, ${\rm Diag}(1, e^{i\alpha}, e^{i\beta})$. The structure of $U_{\rm PMNS}$ matrix can be formed using three rotation matrices with corresponding rotating angles $\theta_{ij}$, and a Dirac CP violating phase $\delta_{CP}$. The PMNS matrix for light neutrino sector is,
	\begin{equation}
	\centering
	U_{\rm PMNS} = 
	\begin{pmatrix} 
	c_{12}c_{13} & s_{12}c_{13} & s_{13} e^{-i\delta_{CP}}  \\
	-s_{12}c_{23} - c_{12}s_{13}s_{23}e^{i\delta_{CP}} & c_{12}c_{23} - s_{12}s_{13}s_{23}e^{i\delta_{CP}} & c_{13}s_{23} \\ s_{12}s_{23} - c_{12}s_{13}c_{23}e^{i\delta_{CP}} & -c_{12}s_{23} - s_{12}s_{13}c_{23}e^{i\delta_{CP}} & c_{13}c_{23}
	\end{pmatrix},                  
	\end{equation}
where $c_{ij} = \cos (\theta_{ij})$, $s_{ij} = \sin (\theta_{ij})$.
After a proper diagonalization, the light and heavy neutrino masses are  generated in the seesaw limit and are given as,
\begin{equation}
m_\nu \simeq f v_L - \frac{v_H^2}{v_R}y_D f^{-1}y_D^T,
\label{eq:nmlr}
\end{equation}
\begin{equation}
M_R =f v_R.
\label{eq:rhmass}
\end{equation}
The expression for light neutrino mass consists both the Majorana and Dirac type of couplings. The first and second term in the right hand side of Eq.({\ref{eq:nmlr}})
are the well known type-II and type-I seesaw mass terms respectively. The so-called
VEV seesaw relation among the various VEVs; $\frac{v_L v_R}{v_H^2}$ can be of ${\mathcal{O}}(1)$ if all the dimensionless couplings in the scalar potential are 
of order one. However, the ratio can take any value between $0$ and $10^{14}$ from 
purely phenomenological point of view.

The effective mass term, $m_{ee}$ ($ee$ element of $M_\nu$) for $0\nu\beta\beta$ decay, in Eq. (\ref{eq:mass}), shows clear dependence on $v_L$ and $v_R$.
So, the procedure followed in section (\ref{sec:dom_types}) for finding the Majorana coupling matrices, makes it a function of $v_R, m_{\text{lightest}}$, mixing angles and Dirac CP phase of $U_{\rm PMNS}$, and Majorana phases, $\alpha$ and $\beta$. The usual mass range for effective electron neutrino ($10^{-5} - 0.1$) eV, suggests the window for a valid $v_R$, which we consider as the new physics scale here.

\section{Interplay of  type-I and type-II mass terms}
\label{sec:dom_types}
In this section we summarize the analytical reconstruction of Majorana  coupling matrix, $f$
in MLRSM as developed in Ref.\cite{Hosteins:2006ja}. For a given hierarchy of light neutrino masses
(normal (NH) or inverted (IH)) and by identifying the criteria to quantify the dominance of type-I or type-II
seesaw, the solutions for the eigenvalues of $f$ and hence the masses of the right-handed neutrino 
are pointed out.

In MLRSM, the masses of the light 
neutrinos result from the interplay of
type-I and type-II mass terms. Using the neutrino mass formula, 
in Eq. (\ref{eq:nmlr}), if
$(f v_L \ll \frac{v_H^2}{v_R}y_D^T f^{-1}y_D)$
the mass becomes,
\begin{equation}
m_\nu \simeq  - \frac{v_H^2}{v_R}y_D f_R^{-1}y_D^T
\label{eq:lnm}.
\end{equation}
This is a case of type-I seesaw dominance and the heavy neutrino masses, $M_i \propto 1/ m_i$. In this paper, the type-I dominance is denoted by $(-)$ sign.
For the condition, 
$(f v_L \gg \frac{v_H^2}{v_R}y_D f^{-1}y_D^T)$
the mass term is,
\begin{equation}
m_\nu \simeq f v_L,
\label{eq:t2d}
\end{equation}
and the heavy neutrino mass term is, $M_i \propto m_i$. This case of type-II dominance is represented by $(+)$ sign. In Eq. (\ref{eq:lnm}), assuming Dirac
type Yukawa coupling is known, $f$ matrix can be reconstructed using low energy neutrino oscillation data. Similarly, Eq. (\ref{eq:t2d}) can be employed to reconstruct the $f$ matrix. In the former case, $f$ is purely
type-I dominated, whereas, in the latter case, it is purely dominated by type-II seesaw. In the following, the analytical method is outlined for
reconstruction of $f$ matrix for one and three generation of leptons considering dominance of both seesaw terms.
\subsubsection{One generation case}
\label{sssec:one_gen}
For one generation of lepton, considering the light neutrino mass formula given in Eq. (\ref{eq:nmlr}),
$m_\nu, y_D$ and $f$ are complex numbers. So Eq. (\ref{eq:nmlr}) is quadratic in $f$. Hence, the two solutions for $f$ are,
\begin{equation}
f_{\pm} = \frac{m_{\nu}}{2 v_L} \left[1 \pm (1 +d)^{1/2}\right],
\end{equation}
where
\begin{equation}
d = \frac{4v_H^2 y_{D}^2}{m_{\nu}^2}\frac{v_L}{v_R}.
\end{equation}
For $|d| \ll 1$, the two solutions are given by,
\begin{equation}
f_{+} \approx \frac{m_\nu}{v_L} +\frac{v_H^2 y_D^2}{v_R m_\nu},  \quad f_{-} \approx - \frac{v_H^2 y_D^2}{v_R m_\nu}.
\label{eq:domn}
\end{equation}
So, one can identify $f_+$ and $f_-$ are type-II and type-I dominant solutions respectively.  Also in Eq. (\ref{eq:domn}), for $v_L \ll m_\nu$ the solutions $f_{+}$ are not perturbative and should be discarded and the only viable solution is 
$f_{-}$.

\subsubsection{Three generation case}
\label{sssec:three_gen}
Within three generations of leptons, the mass expression for light neutrinos given in Eq. (\ref{eq:nmlr}),
is an relation among matrices, $m_\nu, y_D$ and $f$, of order ($3\times 3$). Assuming the Dirac Yukawa matrices to be symmetric and invertible, Eq. (\ref{eq:nmlr}) can be expressed in terms of new parameters as \cite{Hosteins:2006ja},
\begin{equation}
N=\lambda X - \rho X^{-1},
\label{eq:rearrng}
\end{equation}
where $\lambda=v_L$ and $\rho=\frac{v_H^2}{v_R}$.
Here N and X have the following forms,
\begin{equation}
N=y_D^{-1/2}m_\nu (y_D^{-1/2})^T, \quad X=y_D^{-1/2}f (y_D^{-1/2})^T.
\end{equation}
Assuming a particular form of $y_D$ as we discuss in section(\ref{sec:calc}), and by reconstructing $m_\nu$ by using experimental data for a particular hierarchy in the mass eigenvalues of the light neutrinos,  the roots
of the matrix $N$; $n_1,n_2,n_3$ can be determined as, 
\begin{equation}
\text{Det}(N-nI)=0.
\end{equation}
For diagonalizing the matrix, $N$ we use the orthogonal matrix $P_N$, $N=P_N \text{Diag}(n_1,n_2,n_3) P_N^T$ where $P_NP_N^T =1$. Using the matrix $P_N$, matrix $X$ can be diagonalized as $X=P_N \text{Diag}(x_1,x_2,x_3) P_N^T$, where $x_i, i=1,2,3$ are the eigenvalues of $X$.
Using the solutions $x_i$ and $n_i$, the Eq. (\ref{eq:rearrng}) can be written as,
\begin{equation}
n_i=\lambda x_i - \rho x_i^{-1}, i = 1,2,3. 
\label{eq:sol}   
\end{equation}
Now the $f$ matrix can be written as,
\begin{equation}
\begin{split}
f & =y_D^{1/2}X(y_D^{1/2})^T \\
& =y_D^{1/2}P_N \text{Diag}(x_1,x_2,x_3) P_N^T(y_D^{1/2})^T.
\label{eq:f-solution}
\end{split}
\end{equation}
Solving Eq. (\ref{eq:sol}), we  get the general solution
\begin{equation}
x_i^{\pm} = \frac{n_i \pm \text{Sign}({\rm Re}(n_i)\sqrt{n_i^2 + 4\lambda\rho})}{2\lambda}.
\label{eq:eigenvalues}
\end{equation} 
Now, the solutions, $x_i^{\pm}$ can be viewed in two limits.
For $4\lambda\rho \ll n_i^2$, the solutions are,
\begin{equation}
x_i^+ \simeq \frac{n_i}{\lambda},  \quad   x_i^- \simeq \frac{-\rho}{n_i}.
\label{eq:xi-lim1}
\end{equation}  
For the other limit $n_i^2 \ll 4\lambda\rho$, the solutions are
\begin{equation}
x_i^{\pm} \simeq \pm \text{Sign}({\rm Re}(n_i))\sqrt{\frac{\rho}{\lambda}}.
\label{eq:xi-lim2}
\end{equation}
The $2^3 = 8$ solutions \cite{Akhmedov:2006yp} are obtained for $f$, in the form; for example, $(-,-,+)$ using $(x_1^-,x_2^-,x_3^+)$, and $(+,+,-)$ for $(x_1^+,x_2^+,x_3^-)$ and so on. The solution  $x_i^+$ signifies the type-II dominance and $x_i^-$ for the type-I dominance.
\section{Neutrinoless double beta decay}
\label{sec:onbb}
Several theoretical frameworks predict Majorana nature of neutrinos. But in reality we have no evidence of any Majorana fermions including light active neutrinos. A very delicate process such as neutrinoless double beta decay shows in its tree level diagram that two light neutrinos  getting annihilated at a vertex, can be a potential hint for Majorana nature of neutrino. The detection of neutrinoless double beta decay which is a lepton-number-violating process would establish this fact. The half-life of this decay process can be calculated as \cite{Kotila:2012zza},
\begin{equation}
(T_{1/2}^{0\nu\beta\beta})^{-1} = \frac{\Gamma_{0\nu\beta\beta}}{{\rm ln}~ 2} = \frac{G^{0\nu}}{m_e^2} \lvert {m_{0\nu\beta\beta}}\lvert^2  \lvert M_{0\nu} \lvert^2.
\label{eq:hl}
\end{equation}

The measurement of the life time of the process puts an upper
bound on the effective electron neutrino mass, $\lvert m_{0\nu\beta\beta} \lvert$. Within the SM, involving three active neutrinos, the effective mass for electron neutrino is expressed as, $\lvert m_{0\nu\beta\beta} \lvert = \lvert \sum U_{ei}^2 m_i \lvert$, where $U$ is the $U_{\rm PMNS}$, the light neutrino mixing matrix.
Within MLRSM, neutrinoless double beta decay diagrams at tree level include light neutrinos, heavy neutrinos,  $W_L$ and $W_R$ gauge bosons and triplet scalars, $\Delta$'s \cite{Chakrabortty:2012mh}, \cite{BhupalDev:2014qbx}. It can act as one experimentally verifiable observation of MLRSM. For each possible diagram the effective terms are proportional to the respective amplitudes, termed as $\eta$'s (dimensionless parameters) \cite{Chakrabortty:2012mh}, \cite{Borah:2016iqd}, \cite{Barry:2013xxa}. In each $\eta$-term, the index $i=1, 2, 3$ is used to refer three generation of light and heavy neutrino mass states.

The decay width for neutrinoless double beta decay in MLRSM is,
\begin{equation}
\frac{\Gamma_{0\nu\beta\beta}}{{\rm ln}~ 2}=G_{01}^{0\nu}(\lvert \mathbf{M}_{\nu}^{0\nu}(\eta_\nu^{LL}+\eta_\nu^{RR}+\eta_{\Delta_L}^{LL})\lvert^2+  \lvert\mathbf{M}_{N}^{0\nu}(\eta_N^{LL}+\eta_N^{RR}+\eta_{\Delta_R}^{RR})\lvert^2  +\lvert\mathbf{M}_{\lambda}^{0\nu}\eta_\lambda+\mathbf{M}_{\eta}^{0\nu}\eta_\eta\lvert^2).
\label{eq:dw}
\end{equation}
Here, $G_{01}^{0\nu}$ is the phase space factor and $\mathbf{M}_{\nu}^{0\nu}$, $\mathbf{M}_{N}^{0\nu}$, $\mathbf{M}_{\lambda}^{0\nu}$ and $\mathbf{M}_{\eta}^{0\nu}$ are nuclear matrix elements. As the whole process will take place inside the nucleus of atoms, these nuclear matrix elements must be accounted in every diagram. Nuclear matrix elements in general depend on the initial and final nuclear states of the participating nuclei. So, it can vary for different elements. The values used for nuclear matrix elements 
of isotopes of ${\rm Ge}$ and ${\rm Xe}$ are listed in table \ref{tab:npv}.

\begin{figure}[htb]
\begin{center}
	\includegraphics[width=.3\textwidth,height=4cm]{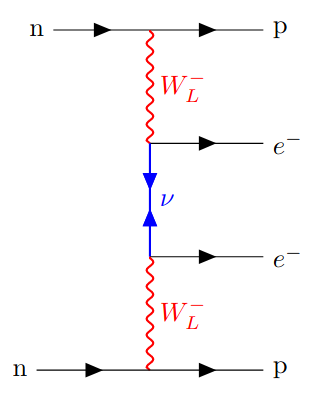}
	\label{fig:lls}
	\hspace{4.0cm}
	\includegraphics[width=.3\textwidth,height=4cm]{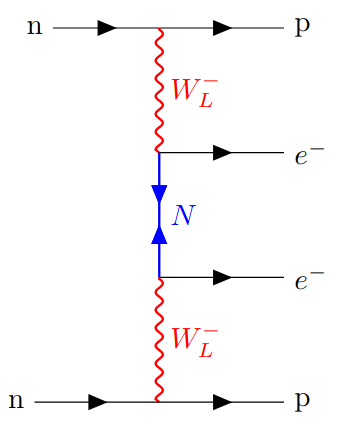}
	\caption{Neutrinoless double beta decay contribution coming from the light (a) and heavy Majorana (b) neutrino with the exchange of $W_L$ gauge bosons.}
	\label{fig:llh}
	\end{center}
\end{figure}

\begin{figure}[htb]
\begin{center}
	\includegraphics[width=.3\textwidth,height=4cm]{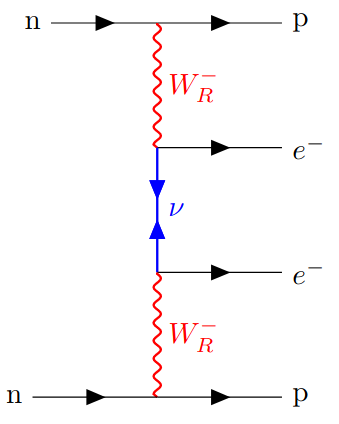}\hspace{0.2cm}
	\hspace{4.0cm}
	\includegraphics[width=.3\textwidth,height=4cm]{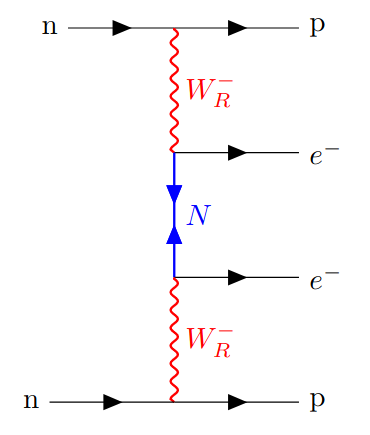}
	\caption{Neutrinoless double beta decay contribution coming from the light (a) and heavy Majorana (b) neutrino with the exchange of $W_R$ gauge bosons.}
	\label{fig:rrh}
	\end{center}
	\end{figure}

\begin{figure}[htb]
	\includegraphics[width=.3\textwidth,height=4cm]{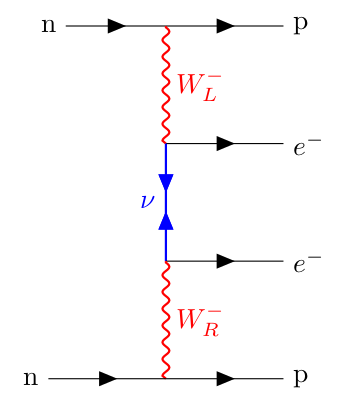}\hspace{0.2cm}
	\hspace{4.0cm}
	\includegraphics[width=.3\textwidth,height=4cm]{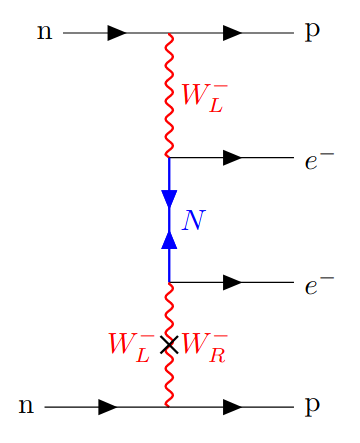}
	\caption{Neutrinoless double beta decay contribution coming from the (a) $\lambda$ diagram (b) neutrino with the mixing of $W_R$ and $W_L$ gauge bosons i.e. the ($\eta$ diagram).}
	\label{fig:mix}
\end{figure}

\begin{figure}[htb]
	\includegraphics[width=.3\textwidth,height=4cm]{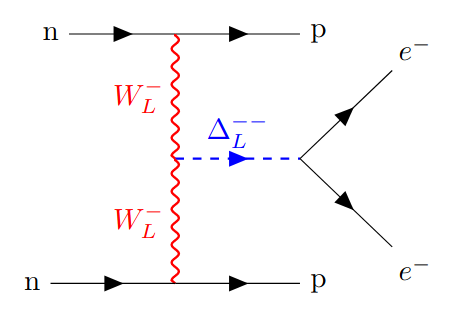}
	\hspace{4.0cm}
	\includegraphics[width=.3\textwidth,height=4cm]{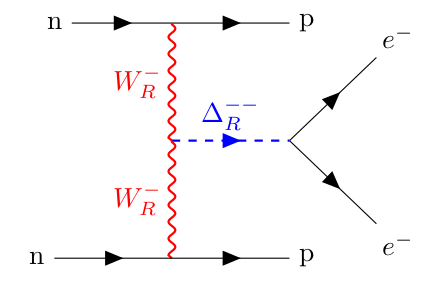}
	\caption{Neutrinoless double beta decay contribution coming from the exchange of $W_R$ and $W_L$ gauge bosons and $\Delta_R$ and $\Delta_L$ bosons ( {\bf See Ref. \cite{Mohapatra:1981pm}}).}
	\label{fig:rrdr}
\end{figure}

The Feynman diagrams for the corresponding decays are shown in figures (\ref{fig:llh} - \ref{fig:rrdr}).

As shown in Fig.(\ref{fig:llh}), the first diagram is the basic tree level diagram for $0\nu\beta\beta$ decay with light active neutrino mediating the process. In the second diagram, the light neutrino is replaced with heavy right-handed neutrino. The dimensionless quantities proportional to their amplitudes are,
\begin{equation}
\eta_\nu^{LL}=\sum_{i} \frac{{U'^2_{L{ei}}} m_i}{m_e},\qquad \eta_N^{LL}= \sum_{i} \frac{T_{L{ei}}^2 m_p}{M_i}.
\end{equation}
In Fig.(\ref{fig:rrh}), the weak gauge bosons, $W_L$ of Fig.(\ref{fig:llh}) are replaced with $W_R$. So, the corresponding contributions are,
\begin{equation}
\eta_\nu^{RR}=\sum_{i} \frac{M_{W_L}^4}{M_{W_R}^4}\frac{S_{ei}^2 m_i}{m_e},  \quad \eta_N^{RR}=\sum_{i}\frac{M_{W_L}^4}{M_{W_R}^4}\frac{{U'^2_{R_{ei}}} m_p}{M_i}. 
\end{equation}
In Fig.(\ref{fig:mix}), the diagrams are called as $\lambda$ and $\eta$ diagrams respectively and the corresponding terms of interest are
\begin{equation}
\eta_{\lambda}=\sum_{i} \frac{M_{W_L}^2}{M_{W_R}^2}U'_{L{ei}} T_{ei}^*, \quad \eta_{\eta}=\sum_{i}\tan(\zeta)U'_{L_{ei}} T_{ei}^*.\\ 
\end{equation}
The diagrams shown in Fig.(\ref{fig:rrdr}) include the doubly charged triplet scalars of both left and right sector. Their contributions are
\begin{equation}
\eta_{\Delta_L}^{LL}=\sum_{i} \frac{{U'^2_{L_{ei}}} m_i m_p}{M_{\Delta_L^{--}}^2},\quad \eta_{\Delta_R}^{RR}=\sum_{i} \frac{M_{W_L}^4}{M_{W_R}^4}\frac{{U'^2_{R_{ei}}} M_i m_p}{M_{\Delta_R^{--}}^2}.
\end{equation}
Now collecting all the terms and comparing with Eq. \ref{eq:hl}, the effective electron neutrino mass $|m_{{\rm \text{eff}}}^{0\nu\beta\beta}|$ can be written as 
\begin{equation}
\lvert{m_{\text{eff}}^{0\nu\beta\beta}}\lvert^2 = m_e^2(\lvert \mathbf{M}_{\nu}^{0\nu}(\eta_\nu^{LL}+\eta_\nu^{RR}+\eta_{\Delta_L}^{LL})\lvert^2 + \lvert\mathbf{M}_{N}^{0\nu}(\eta_N^{LL}+\eta_N^{RR}+\eta_{\Delta_R}^{RR})\lvert^2 +\lvert{\mathbf{M}_{\lambda}^{0\nu}\eta_\lambda+\mathbf{M}_{\eta}^{0\nu}\eta_\eta}\lvert^2).
\label{eq:meff}
\end{equation}
Here the nuclear matrix elements are absorbed within the effective mass expression.
As the expression for $|m_{\text{eff}}^{0\nu\beta\beta}|$ depends upon the masses of right-handed neutrinos, the nature of interplay of type-I and II dominance patterns of Majorana type Yukawa coupling matrices can be studied along with the new physics scale $v_R$.  
\section{Numerical analysis and discussions}
\label{sec:calc}
The effective mass for electron neutrino in $0\nu\beta\beta$ process, shown in Eq. (\ref{eq:meff}), is a function of numerous parameters, like $m_1, m_2, m_3, M_1, M_2, M_3, M_{\Delta_L}, M_{\Delta_R}, M_{W_L}, M_{W_R}$, Majorana phases $( \alpha, \beta)$, Dirac phase ($\delta_{CP}$) and light neutrino mixing angles ($\theta_{12}, \theta_{13}, \theta_{23}$). But such a huge parameter space can be reduced using the neutrino oscillation data. The three small neutrino masses are written in terms of the smallest neutrino mass ($m_1$ for NH and $m_3$ for IH). The masses for right-handed neutrinos, derived using Eq. (\ref{eq:rhmass}) are the functions of $v_R$. The  mass of $W_R$ boson is also proportional to $v_R$. For the remaining terms corresponding to triplet scalars, we have taken $M_{\Delta_L}^{++} = M_{\Delta_R}^{++}$, following the MLRSM and kept it of the order of heaviest right-handed neutrino because the mass for doubly charged particles is also proportional to $v_R$ \cite{Datta:1999nc} (considering the coupling constants of ${\mathcal{O}}(1)$). So, the effective mass becomes a function of $m_1(m_3)$ for NH(IH), $v_R$, $\theta_{ij}$'s, $\delta_{\rm CP}$ and phases, $\alpha$ and $\beta$. The study has been performed for two values of the lightest neutrino mass, $0.01$ eV and $0.001$ eV. For each dominance pattern we have conducted the study by varying the Majorana phases within 0 to $2\pi$. The values for the phase factor and nuclear matrix elements used in analysis are in the following table \ref{tab:npv}.
\begin{table*}[htb]
	\centering
	\setlength{\tabcolsep}{8pt} 
	\renewcommand{\arraystretch}{1.2}
	\begin{center}
		\begin{tabular}{||c |c| c| c| c| c||} 
			\hline
			Isotope & $G_{01}^{0\nu}(yr^{-1})$ & $\mathbf{M}_{\nu}^{0\nu}$ & $\mathbf{M}_{N}^{0\nu}$ & $\mathbf{M}_{\lambda}^{0\nu}$ & $\mathbf{M}_{\eta}^{0\nu}$ \\ [1.0ex] 
			\hline\hline
			Ge-76 & $5.77\times10^{-15}$ & 2.58-6.64 & 233-412 & 1.75-3.76 & 235-637  \\ 
			\hline
			Xe-136 & $3.56\times10^{-14}$ & 1.57-3.85 & 164-172 & 1.92-2.49 & 370-419  \\ 
			\hline
		\end{tabular}
		\caption{Values taken \cite{Kotila:2012zza} for the phase factor and nuclear matrix elements used in calculation. }
		\label{tab:npv}
	\end{center}
\end{table*}
The recent values of these oscillation parameters are given in the following table \ref{tab:tab3}  \cite{Esteban:2020cvm}.
\begin{table*}[htb]
	\centering
	\setlength{\tabcolsep}{8pt} 
	\renewcommand{\arraystretch}{1.2}
	\begin{center}
		\begin{tabular}{|c|c|c|c|c|} 
			\hline
			Parameters & NO(1$\sigma$) & IO(1$\sigma$) & NO(3$\sigma$) & IO(3$\sigma$) \\[1.0ex] 
			\hline
			$\theta_{12}^{\circ}$ & $33.44^{+0.78}_{-0.75}$ & $33.45^{+0.78}_{-0.75}$ & 31.27 - 
			5.86 & 31.27 - 35.87  \\ 
			\hline
			$\theta_{23}^{\circ}$ & $49.2^{+1.1}_{-1.4}$ & $49.5^{+1.0}_{-1.2}$ & 39.6 - 51.8 & 39.9 - 52.0  \\ 
			\hline
			$\theta_{13}^{\circ}$ & $8.57^{+0.13}_{-0.12}$ & $8.61^{+0.12}_{-0.12}$ & 8.20 - 8.97 & 8.24 - 8.98  \\ 
			\hline
			$\delta_{CP}^{\circ}$ & $194^{+51}_{-25}$ & $287^{+27}_{-32}$ & 107 - 403 & 192 - 360  \\ 
			\hline
			$\frac{\Delta m_{21}^2}{10^{-5} eV^2}$ & $7.42^{+0.21}_{-0.20}$ & $7.42^{+0.21}_{-0.20}$ & 6.82 - 8.04 & 6.82 - 8.04  \\ 
			\hline
			$\frac{\Delta m_{3l}^2}{10^{-3} eV^2}$ & $+2.514^{+0.028}_{-0.028}$ &$-2.497^{+0.028}_{-0.028}$ & +2.431 - +2.598 & -2.583 - -2.412  \\ 
			\hline
		\end{tabular}
		\caption{Current values of the neutrino oscillation parameters, taken from \cite{Esteban:2020cvm}. }
		\label{tab:tab3}
	\end{center}
\end{table*}

In order to reconstruct the Majorana coupling matrix $f$, it requires the inputs as $y_D$ and $m_\nu$ at the seesaw scale. Here we adopt a top to bottom approach 
that constructs a viable left-right symmetric theory from a small and controllable 
set of inputs at high scale. The requirement that $f_L = f_R$ and Yukawa coupling matrices are symmetric, arises naturally not only in broad class of left-right symmetric models but also in $SO(10)$ grand unified theory (GUT). $SO(10)$ GUTs have multiple possible breaking patterns to the SM gauge group with intermediate left-right symmetric
model is one of the breaking chains. Starting with the fields at the highest $SO(10)$ scale, one can obtain the fields at the consecutive steps by decomposing representations
of those fields till the SM scale is reached \cite{Deppisch:2017xhv}. The richness of $SO(10)$ GUTs lies in extending the scalar sector with additional exotic scalars fields by providing scopes to use experimental probes to constrain the solutions. Here we consider a class of supersymmetric model with a pair of $126 + \overline{126}$ and two $10$-dimensional Higgs representations \cite{Hosteins:2006ja}. The $\overline{126}$ contains a right-handed triplet $\Delta^c$ with $(1,1,3,-2)$ as well as a left-handed triplet $\Delta$ with $(1,3,1,2)$. Also one $54$-dimensional Higgs representation is added to realize type-II seesaw contribution to the light neutrino mass. A case study based on a class of supersymmetric $SO(10)$ GUT is given in \ref{sec:so10}.
There are a few assumptions made in the calculation. (a) The corrections to the wrong mass relation (Eq.(\ref{eq:gut-mass})) were neglected on the qualitative basis that the mass spectra of the right-handed neutrinos  are detected by the strong hierarchical structure of the Dirac mass matrix and would not be spoiled by the corrections to the basic SO(10) mass relations. 
(b) The gauge breaking aspects of the model including doublet-triplet splitting and gauge symmetry breaking aspects of the model are not taken into account.

In our case the elements of $y_D$  are defined at the GUT scale and are taken to be up-type quark couplings $(y_u,y_c,y_t)$ following the models inspired by grand Unification.  In a basis of the $16$ matter representation in which charged lepton and down-type quark mass matrices are diagonal,
the form for $y_D$ can be constructed as,
\begin{equation}
y_D = U_q^T \hat{y} U_q, \quad U_q = P_u V_{CKM} P_d, \quad \hat{y} = \text{Diag}(y_u,y_c,y_t).
\label{eq:gut-scale}
\end{equation}
Here $P_u$ and $P_d$ are the diagonal matrices which include the phases, ($\phi_u, \phi_c, \phi_t$) and ($\phi_d,\phi_s,  \phi_b$) of the up- and down-type quark fields respectively. Since $y_D$ is weakly renormalizable between GUT and seesaw scales, Eq. (\ref{eq:gut-scale}) is assumed to hold at seesaw scale.
The Dirac Yukawa coupling values for up, charm and top quarks are taken to be $4.2 \times 10^{-6}, 1.75 \times 10^{-3}$ and $0.7$ respectively at GUT scale \cite{Hosteins:2006ja}. The values for Yukawa couplings are then RGE evolved \cite{Rothstein:1990qx} and the help of Mathematica package SARAH \cite{Staub:2013tta} has been taken in this regard. 

\begin{figure*}[h!]
	\centering
	\begin{subfigure}[b]{0.45\textwidth}
		\centering
		\includegraphics[width=\textwidth]{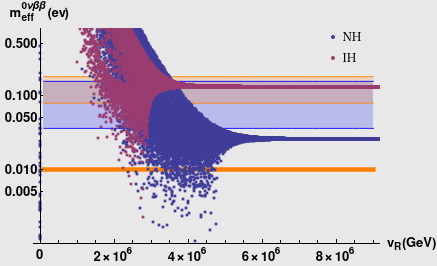}
		\caption*{$(+ + +)$}
	\end{subfigure}
	\hfill
	\begin{subfigure}[b]{0.45\textwidth}
		\centering
		\includegraphics[width=\textwidth]{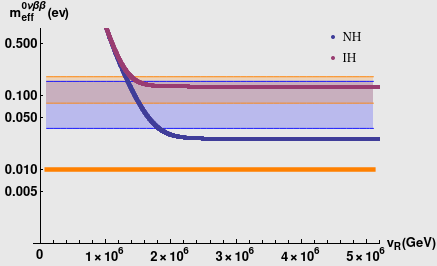}
		\caption*{$(+ - +)$}
	\end{subfigure}
	\hfill
	\begin{subfigure}[b]{0.45\textwidth}
		\centering
		\includegraphics[width=\textwidth]{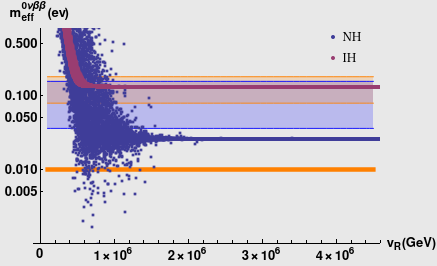}
		\caption*{$(+ + -)$}
	\end{subfigure}
	\hfill
	\begin{subfigure}[b]{0.45\textwidth}
		\centering
		\includegraphics[width=\textwidth]{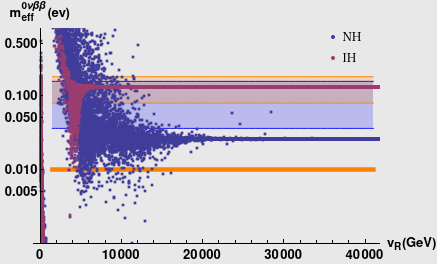}
		\caption*{$(+ - -)$}
	\end{subfigure}
	\hfill
	\begin{subfigure}[b]{0.45\textwidth}
		\centering
		\includegraphics[width=\textwidth]{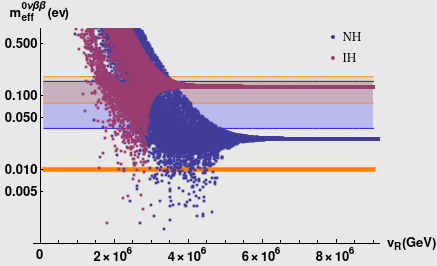}
		\caption*{$(- + +)$}
	\end{subfigure}
	\hfill
	\begin{subfigure}[b]{0.45\textwidth}
		\centering
		\includegraphics[width=\textwidth]{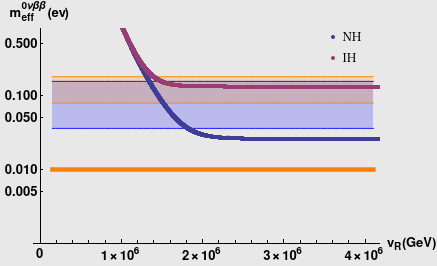}
		\caption*{$(- - +)$}
	\end{subfigure}
	\hfill
	\begin{subfigure}[b]{0.45\textwidth}
		\centering
		\includegraphics[width=\textwidth]{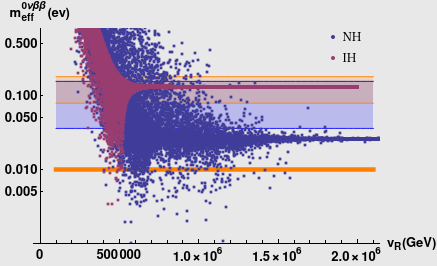}
		\caption*{$(- + -)$}
	\end{subfigure}
	\hfill
	\begin{subfigure}[b]{0.45\textwidth}
		\centering
		\includegraphics[width=\textwidth]{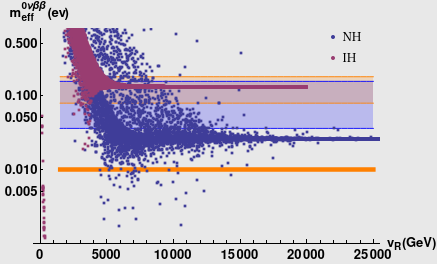}
		\caption*{$(- - -)$}
	\end{subfigure}
	\caption{The figures show combined contribution of active neutrinos and new physics contributions in MLRSM in $|m_{\text{eff}}^{0\nu\beta\beta}|$,
		with variation of $v_R$ as the new physics scale for different dominance 
		patterns in $f$ and hence in the right-handed neutrino masses. The Majorana phases are  varied within $[0,2\pi]$ and $m_{\rm lightest} = 0.01$ eV. Violet and magenta dots represent the results for NH and IH. The limits for effective mass of $0\nu\beta\beta$ decay are given by GERDA as $<(79-180)$ meV \cite{GERDA:2020xhi} (yellow) and KamLand-Zen it is $<(36-156)$ meV \cite{KamLAND-Zen:2022tow} (blue).  The orange lines in all the figures correspond to future sensitivity of $|m_{\text{eff}}^{0\nu\beta\beta}|$.}
	\label{fig:m01}
\end{figure*}

\begin{figure*}[htb]
	\centering
	\begin{subfigure}[b]{0.45\textwidth}
		\centering
		\includegraphics[width=\textwidth]{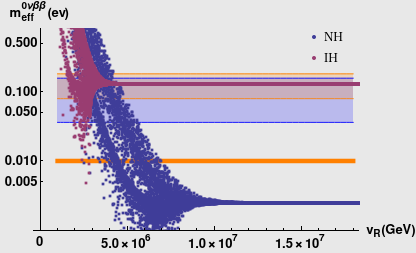}
		\caption*{$(+ + +)$}
	\end{subfigure}
	\hfill
	\begin{subfigure}[b]{0.45\textwidth}
		\centering
		\includegraphics[width=\textwidth]{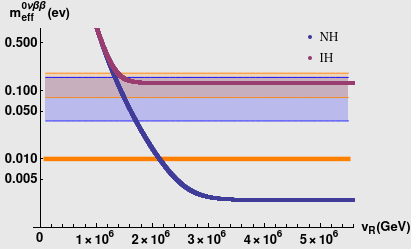}
		\caption*{$(+ - +)$}
	\end{subfigure}
	\hfill
	\begin{subfigure}[b]{0.45\textwidth}
		\centering
		\includegraphics[width=\textwidth]{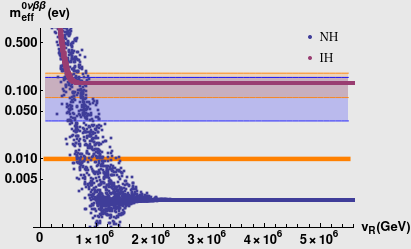}
		\caption*{$(+ + -)$}
	\end{subfigure}
	\hfill
	\begin{subfigure}[b]{0.45\textwidth}
		\centering
		\includegraphics[width=\textwidth]{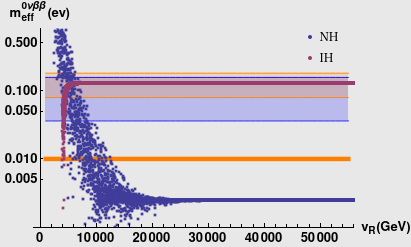}
		\caption*{$(+ - -)$}
	\end{subfigure}
	\hfill
	\begin{subfigure}[b]{0.45\textwidth}
		\centering
		\includegraphics[width=\textwidth]{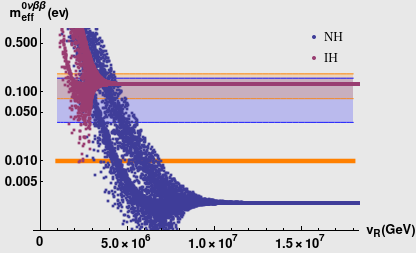}
		\caption*{$(- + +)$}
	\end{subfigure}
	\hfill
	\begin{subfigure}[b]{0.45\textwidth}
		\centering
		\includegraphics[width=\textwidth]{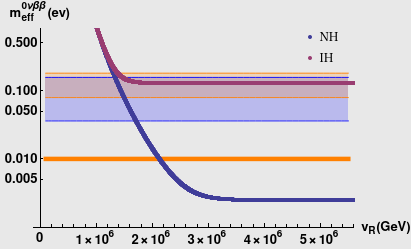}
		\caption*{$(- - +)$}
	\end{subfigure}
	\hfill
	\begin{subfigure}[b]{0.45\textwidth}
		\centering
		\includegraphics[width=\textwidth]{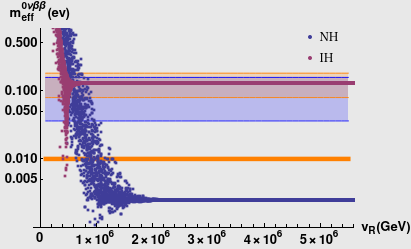}
		\caption*{$(- + -)$}
	\end{subfigure}
	\hfill
	\begin{subfigure}[b]{0.45\textwidth}
		\centering
		\includegraphics[width=\textwidth]{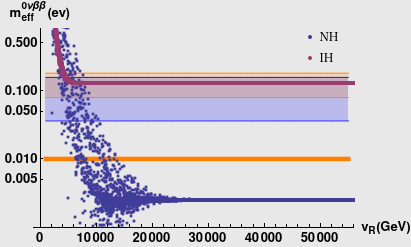}
		\caption*{$(- - -)$}
	\end{subfigure}
	\caption{The figures show combined contribution of active neutrinos and new physics contributions in MLRSM in $|m_{\text{eff}}^{0\nu\beta\beta}|$,
		with variation of $v_R$ as the new physics scale for different dominance 
		patterns in $f$ and hence in the right-handed neutrino masses. The Majorana phases are  varied within $[0,2\pi]$ and $m_{\rm lightest} = 0.001$ eV. Violet and magenta dots represent the results for NH and IH. The limits for effective mass of $0\nu\beta\beta$ decay are given by GERDA as $<(79-180)$ meV \cite{GERDA:2020xhi} (yellow) and KamLand-Zen it is $<(36-156)$ meV \cite{KamLAND-Zen:2022tow} (blue).  The orange lines in all the figures correspond to future sensitivity of $|m_{\text{eff}}^{0\nu\beta\beta}|$.}
	\label{fig:m001}
\end{figure*}

Working in the same basis, the light neutrino mass matrix is given as
\begin{equation}
 M_\nu = U_l^* \hat{M}_\nu U_l^\dagger, \quad U_l = P_e U_{\rm PMNS} P_\nu, \quad
 \hat{M}_\nu = {\rm Diag}(m_1, m_2, m_3),
 \label{eq:numass-gut}
 \end{equation}
where $m_1, m_2$ and $m_3$ are the light neutrino masses at GUT scale. $P_\nu$ is a 
diagonal phase matrix associated with the Majorana nature of the light neutrinos and 
$P_e$ is analogous to the high-energy phases as in $P_u$ and $P_d$.
Here we have not taken the renormalization group evolution of neutrino mass matrix into account, as the result is only
$m_\nu (\rm high ~scale) \sim 1.3~ m_\nu$(low scale) \cite{Barbieri:1999ma,Antusch:2003kp, Giudice:2003jh}. The eigenvalues of the 
Majorana coupling matrix and hence right-handed neutrino masses are reconstructed using the recipe given in section
(\ref{sec:dom_types}). 
In our calculation
we have used equations (\ref{eq:f-solution}) and (\ref{eq:eigenvalues}), to reconstruct $f$ and hence the right-handed 
neutrino mass (Eq.(\ref{eq:rhmass})), which are further used in Eq.(\ref{eq:meff}).
As an example, the right-handed neutrino masses as a function of  
$v_R$ are shown in Fig.(\ref{fig:RH-mass}) in \ref{sec:RH-mass}.
Subsequently, they are used to calculate $|m_{\text{eff}}^{0\nu\beta\beta}|$ as a function of $v_R$. 
In getting the limits on $v_R$ we work with Eq.(\ref{eq:meff}). Apart from other parameters of the theory the effective mass is a function of the masses of the right-handed neutrinos, $M_i$. The $M_i$'s are related to the Majorana coupling matrix $f$ as $M = v_R f$. Thus the effective mass becomes a function of $v_R$ through $M_i$'s and the diagonalizing matrix $U_R$.
As shown in figures (\ref{fig:m01}) and (\ref{fig:m001}), using the bound on the effective mass from present 
and future searches we extract the limits on $v_R$ for $8$ different solutions of 
$f$.

For each  dominance pattern, the variation of $|m_{\text{eff}}^{0\nu\beta\beta}|$ against $v_R$ is observed. For the values of the smallest neutrino mass, $0.01$ eV and $0.001$ eV the results are depicted in figures (\ref{fig:m01}) and (\ref{fig:m001}) respectively. The limits for effective mass of neutrinoless double beta decay are given by GERDA as $<(79-180)$ meV \cite{GERDA:2020xhi} and KamLand-Zen as $<(36-156)$ meV \cite{KamLAND-Zen:2022tow}. 
The yellow and blue regions in the figures (\ref{fig:m01}) and (\ref{fig:m001}) are the upper limit on $|m_{\text{eff}}^{0\nu\beta\beta}|$ from GERDA and KamLand-Zen collaborations respectively. The orange lines in all the figures correspond
to future sensitivity of $|m_{\text{eff}}^{0\nu\beta\beta}|$, in searches of
$0\nu\beta\beta$ decay process. The violet and magenta points are the values of  $|m_{\text{eff}}^{0\nu\beta\beta}|$, obtained numerically, for NH and IH
respectively.
\begin{table*}[htb]
	\setlength{\tabcolsep}{6pt}
	\renewcommand{\arraystretch}{0.7}
	\begin{center}
		\begin{tabular}{||c|c|c|c|c|c||} 
			\hline
			\hline
			\multicolumn{2}{|c|}{}&  \multicolumn{2}{|c|}{($m_{\text{lightest}}$ = 0.01 eV)} & \multicolumn{2}{|c|}{($m_{\text{lightest}}$ = 0.001 eV)} \\			\hline
			Dominance & Order & Lower limit for $v_R$ & Saturation value of $v_R$ & Lower limit for $v_R$ & Saturation value of $v_R$ \\ [1.0ex] 
			patterns&  & (GeV) & (GeV) & (GeV) & (GeV) \\
			\hline
			\hline
			{$+ + +$} & NH & $2.2\times10^6$ & $6.87\times10^6$  & $3\times10^6$ & $1.35\times10^7$\\ 
			& IH & $1.5\times 10^6$ & $5.14\times 10^6$ & $1.2\times 10^6$ & $5\times 10^6$\\ 
			\hline
			{$- + +$} & NH & $2.2\times10^6$ & $6.87\times10^6$ & $3\times10^6$ & $1.35\times10^7$\\ 
			& IH & $1.5\times 10^6$ & $5.14\times 10^6$ & $1.2\times 10^6$ & $5\times 10^6$\\ 
			\hline
			{$+ - +$} & NH & $1.30\times10^6$ & Saturated & $1.5\times 10^6$ & Saturated\\ 
			& IH & $1.38\times 10^6$ & Saturated & $1.68\times 10^6$ & Saturated\\
			\hline
			{$- - +$} & NH & $1.30\times10^6$ & Saturated & $1.5\times 10^6$ & Saturated\\ 
			& IH & $1.38\times 10^6$ & Saturated & $1.68\times 10^6$ & Saturated\\
			\hline
			{$+ + -$} & NH & $3.6\times10^5$ & $2.06\times10^6$  & $4\times10^5$ & $2.5\times10^5$\\ 
			& IH & $4.5\times 10^5$ & $0.97\times 10^6$ & $4\times10^5$ & $9\times10^6$ \\
			\hline
			{$- + -$} & NH & $4\times10^5$ & $2.38\times10^6$ & $4.5\times10^5$ & $2.5\times10^6$\\ 
			& IH & $2.7\times 10^5$ & $0.915\times 10^6$ & $3\times10^5$ & $7.5\times10^5$\\
			\hline
			{$+ - -$} & NH & $2.2\times10^3$ & $3\times10^4$ & $3000$ & $3\times 10^4$\\ 
			& IH & $2.5\times 10^3$ & $9297$ & $4000$ & $6000$\\
			\hline
			{$- - -$} & NH & $3\times10^3$ & $2.7\times10^4$ &$3000$ & $2.5\times 10^4$\\ 
			& IH & $2.5\times 10^3$ & 9577 & $4200$ & $6000$\\
			\hline
		\end{tabular}
		\caption{ Table showing the lower limit  and saturation value of $v_R$, saturating the experimental bound on $|m_{\text{eff}}^{0\nu\beta\beta}|$ for  $m_{\text{lightest}}$ taken as $0.01$ eV and $0.001$ eV for different 
			dominance patterns.}
		\label{tab:vR-values}
	\end{center}
\end{table*}

The combined contribution of active neutrinos and new physics contribution in the MLRSM, to $0\nu \beta\beta$ decay process, saturating the experimental bound shows a narrow range of values of $v_R$. 
The details on the lower limit on $v_R$ and the values at which it saturates in contributing to $|m_{\text{eff}}^{0\nu\beta\beta}|$ are the  shown in 
table (\ref{tab:vR-values}).
The different values of lower limit on $v_R$ range from $\sim (10^{3}- 10^{6})$ GeV for different dominance patterns. For each dominance pattern the range of $v_R$ between the lower limit and the
corresponding saturation value is very narrow. For example, for $(+++)$ pattern, the lower limit on 
$v_R$ and the corresponding value of saturation, the range is from $(2.2 - 6.87) \times 10^{6}$ GeV. 
For ($+ - -$) and ($- - -$) dominance patterns, the set of plots in figures (\ref{fig:m01}) and (\ref{fig:m001}) show, for the range of $v_R$ within 
$10$ TeV, $|m_{\text{eff}}^{0\nu\beta\beta}|$ can saturate the experimental bound.
The suggested energy range  has a potential to be probed in the ongoing experiments as a hint of new physics.
The range beyond the predicted values of $v_R$, can be divided into two regimes:  (a) below the lower limit of $v_R$, (b) above saturation point. Below the lower limit, the parameter space is outside the 
laboratory  bounds of $|m_{\text{eff}}^{0\nu\beta\beta}|$. Above the saturation point, the value of   $|m_{\text{eff}}^{0\nu\beta\beta}|$ is saturated and 
is unaffected by the variation of the parameters of the theory. 
This also indicates there is a lower limit on $v_R$ for which the combined contribution of light neutrinos and new physics contributions in left-right symmetric model contribute to the decay. By comparing figures (\ref{fig:m01}) and (\ref{fig:m001}) and keeping future sensitivity of searches
in mind, it offers more parameter space for the lower values of $m_{\rm lightest}$ 
for most of the dominance patterns.

The results can further be interpreted by comparing the combined contribution as explained above by making a comparison with the standard three active neutrino picture, particular to $0\nu\beta\beta$ decay predictions. In standard three light neutrino picture, where light neutrinos are mediating the $0\nu\beta\beta$ decay process \cite{Barabash:2018fun},
for NH, $|m_{\text{eff}}^{0\nu\beta\beta}| \approx 0 - 30$ meV depending upon different values of $m_{\rm lightest}$.
For $m_{\rm lightest} \leq 10^{-3}$ eV, $|m_{\text{eff}}^{0\nu\beta\beta}| \approx 1 - 4$ meV. For $m_{\rm lightest}$ in the range $(1 - 10)\times 10^{-3}$ eV, there is a possibility of strong cancellation in the  $|m_{\text{eff}}^{0\nu\beta\beta}|$ due to the Majorana phases. For IH, the effective mass $|m_{\text{eff}}^{0\nu\beta\beta}| \approx 10 - 50$ meV for all allowed values of lightest mass $m_{\rm lightest}$.
If $0\nu \beta \beta$ decay is detected, it would be possible to make definitive predictions on the value of
$|m_{\text{eff}}^{0\nu\beta\beta}|$. 
Global analysis of neutrino oscillation data prefers
that  NH is more preferable over IH \cite{Capozzi:2017ipn, Esteban:2020cvm}  at $3.5 \sigma$ level \cite{DeSalas:2018rby}. Assuming neutrinos are Majorana particles, a limit on 
$|m_{\text{eff}}^{0\nu\beta\beta}|$ below $14$ meV would rule out the IH scheme.
The future ton-scale experiments like CUPID \cite{CUPID:2019imh,Armengaud:2019loe}, LEGEND \cite{LEGEND:2017cdu}, and nEXO \cite{nEXO:2017nam,nEXO:2018ylp} are going to probe  $|m_{\text{eff}}^{0\nu\beta\beta}| \le 0.01$ eV.
If $|m_{\text{eff}}^{0\nu\beta\beta}|$ is not registered in the above stated ranges, the presence of new physics contributions other than the light neutrinos can
register very low values of $|m_{\text{eff}}^{0\nu\beta\beta}|$.

Our result is relevant with the statement that NH is more favorable than IH  \cite{DeSalas:2018rby}. As can be seen in Fig.(\ref{fig:m01}),  for $m_{\rm lightest} =0.01$ eV, the future sensitivity of $|m_{\text{eff}}^{0\nu\beta\beta}|$ ($\lesssim 0.01$ eV) will
rule out the parameter space of IH for the mixed solutions $(+-+), (++-),(+--)$ and $(--+)$. For the similar sensitivity of $|m_{\text{eff}}^{0\nu\beta\beta}|$, IH will be ruled out for all solutions for $m_{\rm lightest} =0.001$. For $m_{\rm lightest} = 0.01$ eV, for the dominance pattern  $(+ - +)$ and  $(- - +)$, it will rule out the parameter space of NH. But for $m_{\rm lightest} = 0.001$ eV, these patterns will become accessible for NH.

%

\section{Conclusion}
\label{sec:concl}
With discovery of a Higgs like particle \cite{Baak:2012kk,Carmi:2012in} at ATLAS \cite{ATLAS:2012yve,ATLAS:2015yey,ATLAS:2016neq} and CMS \cite{CMS:2012qbp,CMS:2014suk}  at Large Hadron Collider, the Dirac Yukawa couplings of the charged leptons with the Higgs scalar can be predicted. Making similar predictions for neutral lepton like neutrino is still not possible for example in models where the SM is extended with right-handed neutrinos to generate the light neutrino mass via type-I seesaw mechanism. In LRSM, where the light neutrino mass is generated via combination of type-I and type-II seesaw route, 
the Dirac Yukawa coupling can be uniquely determined as proposed by 
the authors of references \cite{Nemevsek:2012iq} and \cite{Senjanovic:2018xtu}. As a result of determining the Dirac Yukawa couplings, one can make predictions for both high and low energy phenomena such as decays of heavy neutrino, electric dipole moments of charged leptons, neutrinoless double beta decay etc. Although in LRSM the light neutrino mass is of both type-I and type-II origin, these kinds of analyses only take type-I contribution in to account. Unlike Dirac type coupling,  Majorana type 
coupling can not be constrained experimentally. Nevertheless, the seesaw formula including
both type-I and type-II contributions can be employed to reconstruct the Majorana
coupling matrix taking certain quantities as input parameters \cite{Akhmedov:2006de}. The study can provide some insight in to the underlying theory at the seesaw scale. Taking experimentally observable data for
light neutrino masses and mixing as input and assuming up-quark type Dirac Yukawa coupling for the neutrinos,
the Majorana coupling (and hence the masses of the right-handed neutrinos) can be expressed numerically as function of the left-right symmetry breaking scale, $v_R$ \cite{Hosteins:2006ja}. Depending upon dominance of type- I and type -II
seesaw terms and interplay among them, total eight solutions for the Majorana coupling are obtained. We have studied the signatures of the eight solutions taking $0\nu\beta\beta$ decay process in to account. 
The effective mass contains  contributions from all the possible diagrams (tree level) in MLRSM. Specifically, the effective mass is a function of the masses of lightest active neutrinos, three heavy neutrinos, triplet scalars,
gauge bosons and the Majorana phases ($\alpha,\beta$).

The combined contribution to $0\nu \beta\beta$ including active neutrinos and new physics contribution in the MLRSM, covers a significant parameter space of  $|m_{\text{eff}}^{0\nu\beta\beta}|~~ vs~~ m_{\rm lightest}$ depending on varying Majorana phases but at a particular value of $v_R$ each profile gets saturated to a single line and continues further. For different dominance patterns the saturation value of $v_R$ is different and is in a range between $\sim (10^{3}- 10^{6})$ GeV. 
 The range beyond the predicted values of $v_R$, can be divided into two regimes:  (a) below the lower limit of $v_R$, (b) above saturation point. Below the lower limit, the parameter space is outside the 
laboratory  bounds of $|m_{\text{eff}}^{0\nu\beta\beta}|$. Above the saturation point, the value of   $|m_{\text{eff}}^{0\nu\beta\beta}|$ is saturated and 
is unaffected by the variation of the parameters of the theory. 
This also indicates there is a lower limit on $v_R$ for which the combined contribution of light neutrinos and new physics contributions in left-right symmetric model contribute to the decay. By comparing figures (\ref{fig:m01}) and (\ref{fig:m001}) and keeping future sensitivity of searches
in mind, it offers more parameter space for the lower values of $m_{\rm lightest}$ 
for most of the dominance patterns.

The future ton-scale experiments like CUPID \cite{CUPID:2019imh,Armengaud:2019loe}, LEGEND \cite{LEGEND:2017cdu}, and Neo \cite{nEXO:2017nam,nEXO:2018ylp} are going to probe  $|m_{\text{eff}}^{0\nu\beta\beta}| \le 0.01$ eV.
Our result is relevant with the statement that NH is more favorable than IH.  For $m_{\rm lightest} =0.01$ eV, for the future sensitivity of $|m_{\text{eff}}^{0\nu\beta\beta}| \le 0.01$ eV will
rule out the parameter space of IH for the mixed solutions $(+-+), (++-),(+--)$ and $(--+)$. 
For $m_{\rm lightest} = 0.01$ eV, for the dominance pattern of  $(+ - +)$ and  $(- - +)$, the parameter space of NH will be ruled out. But for $m_{\rm lightest} = 0.001$ eV, for these patterns, it will become accessible for NH. For $m_{\rm lightest} = 0.001$ eV, it shows that the parameter space of IH will be completely
out of future sensitivity of search $0\nu \beta\beta$ decay.
The data from Cosmology could possibly impose a lower limit
on the lightest neutrino mass.  
As a  general observation, by lowering the value of $m_{\rm lightest}$ from $0.01$ eV to $0.001$ eV
the effective mass is lowered towards the future sensitivity region.
In future even if combined cosmological observations \cite{Brinckmann:2018owf} push total sum of light neutrino mass below $0.1$ eV, there
will be some parameter space available to probe for new physics giving
dominant contribution to the $0\nu\beta\beta$ decay process.

\section*{Acknowledgements}
We are grateful to Prof. Evgeny K. Akhmedov for the insightful discussions throughout the work. V.B would like to thank MHRD, Govt. of India for financial assistance.

\appendix
\section{A case study}
\label{sec:so10}
With the set of
Higgs representations as mentioned in section(\ref{sec:calc}), the general Yukawa interaction terms can be written as
\begin{equation}
 \overline{16}_F ~\left( Y^{(1)} 10_1 + Y^{(2)} 10_2 + f ~\overline{126} \right) ~16_F,
\label{eq:yuk-gut}
 \end{equation}
where $16_F$ represents the $16$-dimensional matter field representation.
$Y^{(1)}, Y^{(2)}$ and $f$ are complex symmetric matrices.
If the fermion masses originate only from the VEVs of the 
scalars in two $10$ representations, not from the 
scalars in $126$ representation, a mass relation
among charged fermions can be realized from Eq. (\ref{eq:yuk-gut}), at the GUT scale;
\begin{equation}
 M_u =M_D, \quad M_d = M_e
 \label{eq:gut-mass}
\end{equation}
The two relations in Eq.(\ref{eq:gut-mass}) are often referred as quark-lepton symmetry.
The second relation in Eq. (\ref{eq:gut-mass}) is not in agreement with experimental data.
It can be corrected by considering contributions from $SU(2)_L$
doublet \cite{Mohapatra:1979nn,Babu:1992ia} or from non-renormalizable operators \cite{Anderson:1993fe}. Although, these corrections
can affect the first relation, and subsequently change the numerical solution for $f$, it does not alter the qualitative features of the present study. The mass spectra of the right-handed neutrinos
are detected by the strong hierarchical structure of the Dirac mass matrix and would not be spoiled by the
corrections to the basic $SO(10)$ mass relations.
The boundary condition for $y_D$ at GUT scale is thus taken according to the 
first relation in Eq. (\ref{eq:gut-mass}) for reconstructing the $8$ solutions of $f$ matrix.

The solutions of $f$ matrix subsequently give rise to right-handed neutrino masses and
mixing, which also depend on the ratio $\lambda/\rho$ and $v_R$.
The type-II seesaw mechanism can be realized by adding a $54$ representation in the Higgs sector in addition to the $126 + \overline{126}$ pair.
The $54$ contains a bitriplet $\tilde{\Delta}$ with $(3,3,1,0)$ and 
each $10$ contains a bidoublet $\varphi$ with $(1,2,2,0)$. The terms relevant for the
type-II seesaw contribution are,
\begin{equation}
 W = \frac{1}{2} f_{ij} L_i L_J \Delta + \mu \varphi \varphi \tilde{\Delta} +
 \tilde{\mu} \Delta \Delta^c \tilde{\Delta}+ \cdots.
\end{equation}
The first term comes from the couplings of the form $16_i~16_j~ \overline{126}$, and 
the second and third terms come from $10 ~ 10~ 54$, $54~ \overline{126}~\overline{126}$ couplings. From the above terms, a VEV for the $SU(2)_L$ triplet is induced, 
$v_L \sim \mu \tilde{\mu} v_u^2 v_R /M^2_{\Delta_L}$. Thus, the ratio $\lambda/\rho
= \sim v_u^2 /(v_L v_R)\sim M^2_{\Delta}/(\mu \tilde{\mu} v_R^2)$.
Depending on the superpotential parameters $\lambda/\rho$ can be larger or smaller than $1$. The scale of $v_R$ which related to the breaking scheme of the $SO(10)$ symmetry is free parameter between the GUT and electroweak scales. 
\section{Masses of the right-handed neutrinos}
\label{sec:RH-mass}
\begin{figure*}[htb]
	\centering
	\begin{subfigure}[b]{0.4\textwidth}
		\centering
		\includegraphics[width=\textwidth]{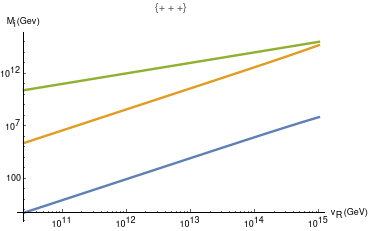}
	\end{subfigure}
	\hfill
	\begin{subfigure}[b]{0.4\textwidth}
		\centering
		\includegraphics[width=\textwidth]{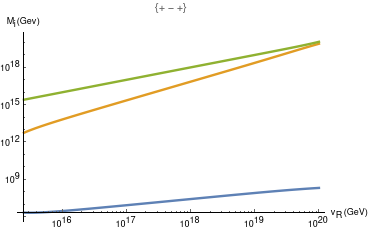}
	\end{subfigure}
	\bigskip
	\begin{subfigure}[b]{0.4\textwidth}
		\centering
		\includegraphics[width=\textwidth]{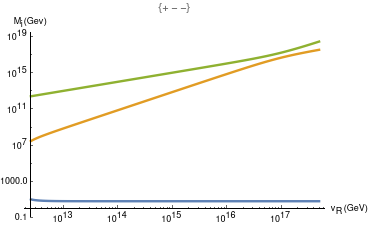}
	\end{subfigure}
	\hfill
	\begin{subfigure}[b]{0.4\textwidth}
		\centering
		\includegraphics[width=\textwidth]{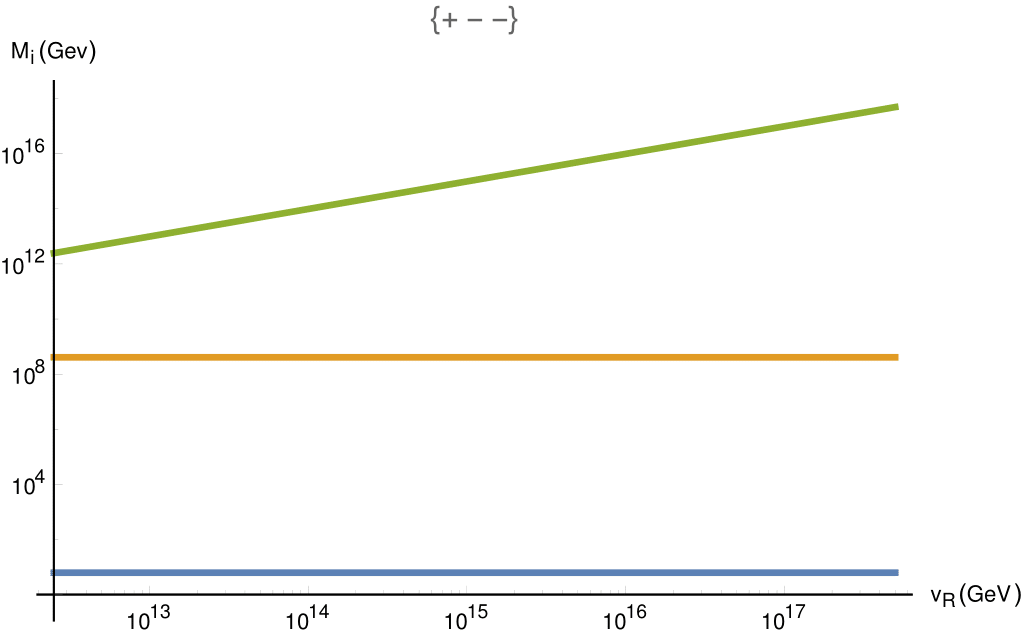}
	\end{subfigure}
	\bigskip
	\begin{subfigure}[b]{0.4\textwidth}
		\centering
		\includegraphics[width=\textwidth]{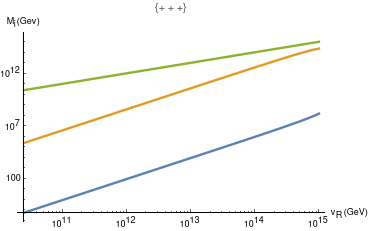}
	\end{subfigure}
	\hfill
	\begin{subfigure}[b]{0.4\textwidth}
		\centering
		\includegraphics[width=\textwidth]{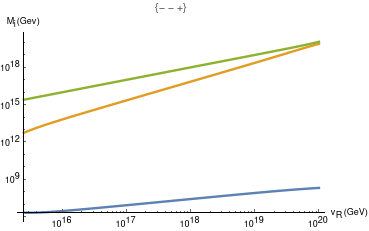}
	\end{subfigure}
	\bigskip
	\begin{subfigure}[b]{0.4\textwidth}
		\centering
		\includegraphics[width=\textwidth]{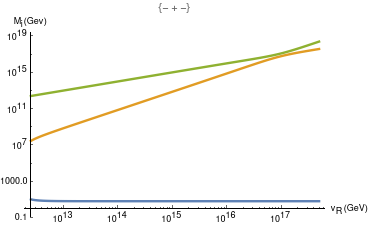}
	\end{subfigure}
	\hfill
	\begin{subfigure}[b]{0.4\textwidth}
		\centering
		\includegraphics[width=\textwidth]{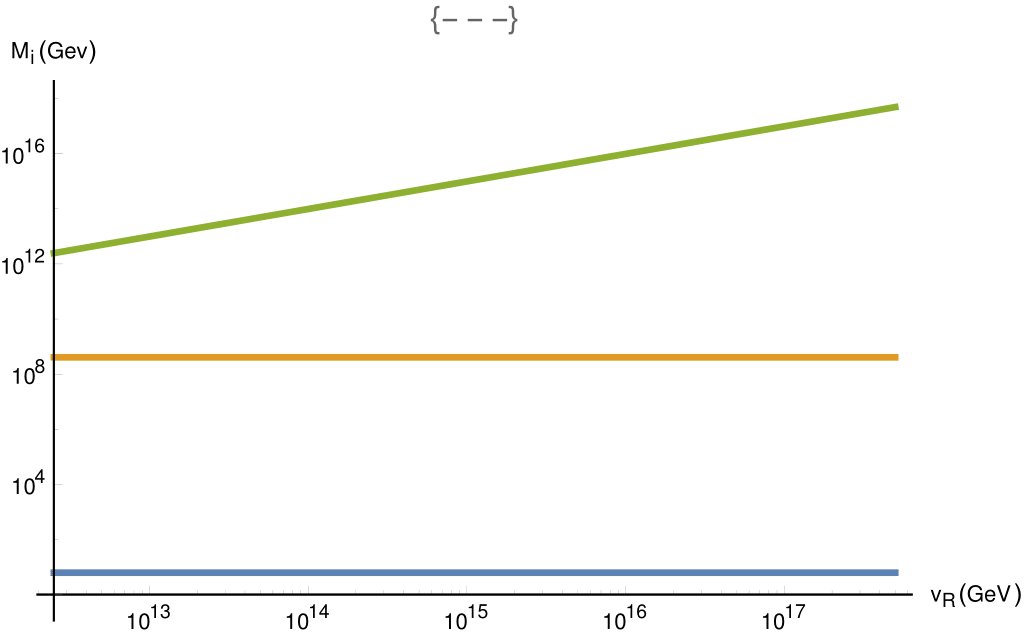}
	\end{subfigure}
	\caption{The figures show the right-handed neutrino masses as 
		a function of $v_R$ for eight different dominance patterns of
		for NH, taking $m_{\rm lightest} = 0.01$ eV.}
	\label{fig:RH-mass}
\end{figure*}

We can get the right-handed neutrino mass spectrum for each dominance pattern as $M_i = v_Rf_i $. As an example for NH, taking $m_{\rm lightest} = 0.01$ eV,
they are shown in Fig.(\ref{fig:RH-mass}) (blue for $M_1$, orange for $M_2$ and  green for $M_3$).

\bibliographystyle{elsarticle-num} 

\begin{thebibliography}{10}
\expandafter\ifx\csname url\endcsname\relax
  \def\url#1{\texttt{#1}}\fi
\expandafter\ifx\csname urlprefix\endcsname\relax\def\urlprefix{URL }\fi
\expandafter\ifx\csname href\endcsname\relax
  \def\href#1#2{#2} \def\path#1{#1}\fi

\bibitem{Giunti:2007ry}
C.~Giunti, C.~W. Kim, {Fundamentals of Neutrino Physics and Astrophysics},
  2007.

\bibitem{Mohapatra:1974gc}
R.~N. Mohapatra, J.~C. Pati, {A Natural Left-Right Symmetry}, Phys. Rev. D 11
  (1975) 2558.
\newblock \href {http://dx.doi.org/10.1103/PhysRevD.11.2558}
  {\path{doi:10.1103/PhysRevD.11.2558}}.

\bibitem{Senjanovic:1975rk}
G.~Senjanovic, R.~N. Mohapatra, {Exact Left-Right Symmetry and Spontaneous
  Violation of Parity}, Phys. Rev. D 12 (1975) 1502.
\newblock \href {http://dx.doi.org/10.1103/PhysRevD.12.1502}
  {\path{doi:10.1103/PhysRevD.12.1502}}.

\bibitem{Senjanovic:1978ev}
G.~Senjanovic, {Spontaneous Breakdown of Parity in a Class of Gauge Theories},
  Nucl. Phys. B 153 (1979) 334--364.
\newblock \href {http://dx.doi.org/10.1016/0550-3213(79)90604-7}
  {\path{doi:10.1016/0550-3213(79)90604-7}}.

\bibitem{Georgi:1974my}
H.~Georgi, {The State of the Art\textemdash{}Gauge Theories}, AIP Conf. Proc.
  23 (1975) 575--582.
\newblock \href {http://dx.doi.org/10.1063/1.2947450}
  {\path{doi:10.1063/1.2947450}}.

\bibitem{Fritzsch:1974nn}
H.~Fritzsch, P.~Minkowski, {Unified Interactions of Leptons and Hadrons},
  Annals Phys. 93 (1975) 193--266.
\newblock \href {http://dx.doi.org/10.1016/0003-4916(75)90211-0}
  {\path{doi:10.1016/0003-4916(75)90211-0}}.

\bibitem{Minkowski:1977sc}
P.~Minkowski, {$\mu \to e\gamma$ at a Rate of One Out of $10^{9}$ Muon
  Decays?}, Phys. Lett. B 67 (1977) 421--428.
\newblock \href {http://dx.doi.org/10.1016/0370-2693(77)90435-X}
  {\path{doi:10.1016/0370-2693(77)90435-X}}.

\bibitem{Ramond:1979py}
P.~Ramond, {The Family Group in Grand Unified Theories}, in: {International
  Symposium on Fundamentals of Quantum Theory and Quantum Field Theory}, 1979.
\newblock \href {http://arxiv.org/abs/hep-ph/9809459}
  {\path{arXiv:hep-ph/9809459}}.

\bibitem{Gell-Mann:1979vob}
M.~Gell-Mann, P.~Ramond, R.~Slansky, {Complex Spinors and Unified Theories},
  Conf. Proc. C 790927 (1979) 315--321.
\newblock \href {http://arxiv.org/abs/1306.4669} {\path{arXiv:1306.4669}}.

\bibitem{Sawada:1979dis}
O.~Sawada, A.~Sugamoto (Eds.), {Proceedings: Workshop on the Unified Theories
  and the Baryon Number in the Universe}: {Tsukuba, Japan, February 13-14,
  1979}, Natl.Lab.High Energy Phys., Tsukuba, Japan, 1979.

\bibitem{Levy:1980ws}
{QUARKS AND LEPTONS. PROCEEDINGS, SUMMER INSTITUTE, CARGESE, FRANCE, JULY 9-29,
  1979}, Vol.~61.
\newblock \href {http://dx.doi.org/10.1007/978-1-4684-7197-7}
  {\path{doi:10.1007/978-1-4684-7197-7}}.

\bibitem{Mohapatra:1979ia}
R.~N. Mohapatra, G.~Senjanovic, {Neutrino Mass and Spontaneous Parity
  Nonconservation}, Phys. Rev. Lett. 44 (1980) 912.
\newblock \href {http://dx.doi.org/10.1103/PhysRevLett.44.912}
  {\path{doi:10.1103/PhysRevLett.44.912}}.

\bibitem{Magg:1980ut}
M.~Magg, C.~Wetterich, {Neutrino Mass Problem and Gauge Hierarchy}, Phys. Lett.
  B 94 (1980) 61--64.
\newblock \href {http://dx.doi.org/10.1016/0370-2693(80)90825-4}
  {\path{doi:10.1016/0370-2693(80)90825-4}}.

\bibitem{Lazarides:1980nt}
G.~Lazarides, Q.~Shafi, C.~Wetterich, {Proton Lifetime and Fermion Masses in an
  SO(10) Model}, Nucl. Phys. B 181 (1981) 287--300.
\newblock \href {http://dx.doi.org/10.1016/0550-3213(81)90354-0}
  {\path{doi:10.1016/0550-3213(81)90354-0}}.

\bibitem{Mohapatra:1980yp}
R.~N. Mohapatra, G.~Senjanovic, {Neutrino Masses and Mixings in Gauge Models
  with Spontaneous Parity Violation}, Phys. Rev. D 23 (1981) 165.
\newblock \href {http://dx.doi.org/10.1103/PhysRevD.23.165}
  {\path{doi:10.1103/PhysRevD.23.165}}.

\bibitem{Schechter:1980gr}
J.~Schechter, J.~W.~F. Valle, {Neutrino Masses in SU(2) x U(1) Theories}, Phys.
  Rev. D 22 (1980) 2227.
\newblock \href {http://dx.doi.org/10.1103/PhysRevD.22.2227}
  {\path{doi:10.1103/PhysRevD.22.2227}}.

\bibitem{Casas:2001sr}
J.~A. Casas, A.~Ibarra, {Oscillating neutrinos and $\mu \to e, \gamma$}, Nucl.
  Phys. B 618 (2001) 171--204.
\newblock \href {http://arxiv.org/abs/hep-ph/0103065}
  {\path{arXiv:hep-ph/0103065}}, \href
  {http://dx.doi.org/10.1016/S0550-3213(01)00475-8}
  {\path{doi:10.1016/S0550-3213(01)00475-8}}.

\bibitem{Nemevsek:2012iq}
M.~Nemevsek, G.~Senjanovic, V.~Tello, {Connecting Dirac and Majorana Neutrino
  Mass Matrices in the Minimal Left-Right Symmetric Model}, Phys. Rev. Lett.
  110~(15) (2013) 151802.
\newblock \href {http://arxiv.org/abs/1211.2837} {\path{arXiv:1211.2837}},
  \href {http://dx.doi.org/10.1103/PhysRevLett.110.151802}
  {\path{doi:10.1103/PhysRevLett.110.151802}}.

\bibitem{Senjanovic:2018xtu}
G.~Senjanovic, V.~Tello, {Disentangling the seesaw mechanism in the minimal
  left-right symmetric model}, Phys. Rev. D 100~(11) (2019) 115031.
\newblock \href {http://arxiv.org/abs/1812.03790} {\path{arXiv:1812.03790}},
  \href {http://dx.doi.org/10.1103/PhysRevD.100.115031}
  {\path{doi:10.1103/PhysRevD.100.115031}}.

\bibitem{Kiers:2022cyc}
J.~Kiers, K.~Kiers, A.~Szynkman, T.~Tarutina, {Disentangling the seesaw
  mechanism in the left-right model: An algorithm for the general case}, Phys.
  Rev. D 107~(7) (2023) 075001.
\newblock \href {http://arxiv.org/abs/2212.14837} {\path{arXiv:2212.14837}},
  \href {http://dx.doi.org/10.1103/PhysRevD.107.075001}
  {\path{doi:10.1103/PhysRevD.107.075001}}.

\bibitem{Keung:1983uu}
W.-Y. Keung, G.~Senjanovic, {Majorana Neutrinos and the Production of the
  Right-handed Charged Gauge Boson}, Phys. Rev. Lett. 50 (1983) 1427.
\newblock \href {http://dx.doi.org/10.1103/PhysRevLett.50.1427}
  {\path{doi:10.1103/PhysRevLett.50.1427}}.

\bibitem{Akhmedov:2006de}
E.~K. Akhmedov, M.~Frigerio, {Interplay of type I and type II seesaw
  contributions to neutrino mass}, JHEP 01 (2007) 043.
\newblock \href {http://arxiv.org/abs/hep-ph/0609046}
  {\path{arXiv:hep-ph/0609046}}, \href
  {http://dx.doi.org/10.1088/1126-6708/2007/01/043}
  {\path{doi:10.1088/1126-6708/2007/01/043}}.

\bibitem{ATLAS:2019isd}
M.~Aaboud, et~al., {Search for a right-handed gauge boson decaying into a
  high-momentum heavy neutrino and a charged lepton in $pp$ collisions with the
  ATLAS detector at $\sqrt{s}=13$ TeV}, Phys. Lett. B 798 (2019) 134942.
\newblock \href {http://arxiv.org/abs/1904.12679} {\path{arXiv:1904.12679}},
  \href {http://dx.doi.org/10.1016/j.physletb.2019.134942}
  {\path{doi:10.1016/j.physletb.2019.134942}}.

\bibitem{Hosteins:2006ja}
P.~Hosteins, S.~Lavignac, C.~A. Savoy, {Quark-Lepton Unification and Eight-Fold
  Ambiguity in the Left-Right Symmetric Seesaw Mechanism}, Nucl. Phys. B 755
  (2006) 137--163.
\newblock \href {http://arxiv.org/abs/hep-ph/0606078}
  {\path{arXiv:hep-ph/0606078}}, \href
  {http://dx.doi.org/10.1016/j.nuclphysb.2006.07.028}
  {\path{doi:10.1016/j.nuclphysb.2006.07.028}}.
\bibitem{Hirsch:1997tr}
  M.~Hirsch, H.~V. Klapdor-Kleingrothaus, Probing physics beyond the standard model with neutrinoless double beta decay, Nucl. Phys. B Proc. Suppl. 52 (1997) 257--262.
\newblock \href {http://dx.doi.org/10.1016/S0920-5632(96)00574-9}
  {\path{doi:10.1016/S0920-5632(96)00574-9}}.  

\bibitem{Dvali:2023snt}
  G.~Dvali, A.~Maiezza, G.~Senjanovic, V.~Tello, Neutrinoless double beta decay: Neutrino mass versus new physics, Phys. Rev. D 108 (2023) 075012.
\newblock \href {http://arxiv.org/abs/2303.17261} {\path{arXiv:2303.17261}},
  \href {http://dx.doi.org/10.1103/PhysRevD.108.075012}
  {\path{doi:10.1103/PhysRevD.108.075012}}.
\bibitem{Pascoli:2003uh}
S.~Pascoli, S.~T. Petcov, W.~Rodejohann, {On the connection of leptogenesis
  with low-energy CP violation and LFV charged lepton decays}, Phys. Rev. D 68
  (2003) 093007.
\newblock \href {http://arxiv.org/abs/hep-ph/0302054}
  {\path{arXiv:hep-ph/0302054}}, \href
  {http://dx.doi.org/10.1103/PhysRevD.68.093007}
  {\path{doi:10.1103/PhysRevD.68.093007}}.

\bibitem{Buccella:2017jkx}
F.~Buccella, M.~Chianese, G.~Mangano, G.~Miele, S.~Morisi, P.~Santorelli, {A
  neutrino mass-mixing sum rule from SO(10) and neutrinoless double beta
  decay}, JHEP 04 (2017) 004.
\newblock \href {http://arxiv.org/abs/1701.00491} {\path{arXiv:1701.00491}},
  \href {http://dx.doi.org/10.1007/JHEP04(2017)004}
  {\path{doi:10.1007/JHEP04(2017)004}}.

\bibitem{Maiezza:2016ybz}
A.~Maiezza, G.~Senjanovi\'c, J.~C. Vasquez, {Higgs sector of the minimal
  left-right symmetric theory}, Phys. Rev. D 95~(9) (2017) 095004.
\newblock \href {http://arxiv.org/abs/1612.09146} {\path{arXiv:1612.09146}},
  \href {http://dx.doi.org/10.1103/PhysRevD.95.095004}
  {\path{doi:10.1103/PhysRevD.95.095004}}.

\bibitem{Abada:2007ux}
A.~Abada, C.~Biggio, F.~Bonnet, M.~B. Gavela, T.~Hambye, {Low energy effects of
  neutrino masses}, JHEP 12 (2007) 061.
\newblock \href {http://arxiv.org/abs/0707.4058} {\path{arXiv:0707.4058}},
  \href {http://dx.doi.org/10.1088/1126-6708/2007/12/061}
  {\path{doi:10.1088/1126-6708/2007/12/061}}.

\bibitem{Deshpande:1990ip}
N.~G. Deshpande, J.~F. Gunion, B.~Kayser, F.~I. Olness, {Left-right symmetric
  electroweak models with triplet Higgs}, Phys. Rev. D 44 (1991) 837--858.
\newblock \href {http://dx.doi.org/10.1103/PhysRevD.44.837}
  {\path{doi:10.1103/PhysRevD.44.837}}.

\bibitem{Grimus:2000vj}
W.~Grimus, L.~Lavoura, {The Seesaw mechanism at arbitrary order: Disentangling
  the small scale from the large scale}, JHEP 11 (2000) 042.
\newblock \href {http://arxiv.org/abs/hep-ph/0008179}
  {\path{arXiv:hep-ph/0008179}}, \href
  {http://dx.doi.org/10.1088/1126-6708/2000/11/042}
  {\path{doi:10.1088/1126-6708/2000/11/042}}.

\bibitem{Chakrabortty:2012mh}
J.~Chakrabortty, H.~Z. Devi, S.~Goswami, S.~Patra, {Neutrinoless double-$\beta$
  decay in TeV scale Left-Right symmetric models}, JHEP 08 (2012) 008.
\newblock \href {http://arxiv.org/abs/1204.2527} {\path{arXiv:1204.2527}},
  \href {http://dx.doi.org/10.1007/JHEP08(2012)008}
  {\path{doi:10.1007/JHEP08(2012)008}}.

\bibitem{Akhmedov:2006yp}
E.~K. Akhmedov, M.~Blennow, T.~Hallgren, T.~Konstandin, T.~Ohlsson, {Stability
  and leptogenesis in the left-right symmetric seesaw mechanism}, JHEP 04
  (2007) 022.
\newblock \href {http://arxiv.org/abs/hep-ph/0612194}
  {\path{arXiv:hep-ph/0612194}}, \href
  {http://dx.doi.org/10.1088/1126-6708/2007/04/022}
  {\path{doi:10.1088/1126-6708/2007/04/022}}.

\bibitem{Kotila:2012zza}
J.~Kotila, F.~Iachello, {Phase space factors for double-$\beta$ decay}, Phys.
  Rev. C 85 (2012) 034316.
\newblock \href {http://arxiv.org/abs/1209.5722} {\path{arXiv:1209.5722}},
  \href {http://dx.doi.org/10.1103/PhysRevC.85.034316}
  {\path{doi:10.1103/PhysRevC.85.034316}}.

\bibitem{BhupalDev:2014qbx}
P.~S. Bhupal~Dev, S.~Goswami, M.~Mitra, {TeV Scale Left-Right Symmetry and
  Large Mixing Effects in Neutrinoless Double Beta Decay}, Phys. Rev. D 91~(11)
  (2015) 113004.
\newblock \href {http://arxiv.org/abs/1405.1399} {\path{arXiv:1405.1399}},
  \href {http://dx.doi.org/10.1103/PhysRevD.91.113004}
  {\path{doi:10.1103/PhysRevD.91.113004}}.

\bibitem{Borah:2016iqd}
D.~Borah, A.~Dasgupta, {Charged lepton flavour violcxmation and neutrinoless
  double beta decay in left-right symmetric models with type I+II seesaw}, JHEP
  07 (2016) 022.
\newblock \href {http://arxiv.org/abs/1606.00378} {\path{arXiv:1606.00378}},
  \href {http://dx.doi.org/10.1007/JHEP07(2016)022}
  {\path{doi:10.1007/JHEP07(2016)022}}.

\bibitem{Barry:2013xxa}
J.~Barry, W.~Rodejohann, {Lepton number and flavour violation in TeV-scale
  left-right symmetric theories with large left-right mixing}, JHEP 09 (2013)
  153.
\newblock \href {http://arxiv.org/abs/1303.6324} {\path{arXiv:1303.6324}},
  \href {http://dx.doi.org/10.1007/JHEP09(2013)153}
  {\path{doi:10.1007/JHEP09(2013)153}}.

\bibitem{Mohapatra:1981pm}
R.~N. Mohapatra, J.~D. Vergados, A new contribution to neutrinoless double beta decay in gauge models, Phys. Rev. Lett. 47 (1981) 1713--1716.
\newblock \href {http://dx.doi.org/10.1103/PhysRevLett.47.1713}
  {\path{doi:10.1103/PhysRevLett.47.1713}}.
  
\bibitem{Datta:1999nc}
A.~Datta, A.~Raychaudhuri, {Mass bounds for triplet scalars of the left-right
  symmetric model and their future detection prospects}, Phys. Rev. D 62 (2000)
  055002.
\newblock \href {http://arxiv.org/abs/hep-ph/9905421}
  {\path{arXiv:hep-ph/9905421}}, \href
  {http://dx.doi.org/10.1103/PhysRevD.62.055002}
  {\path{doi:10.1103/PhysRevD.62.055002}}.

\bibitem{Esteban:2020cvm}
I.~Esteban, M.~C. Gonzalez-Garcia, M.~Maltoni, T.~Schwetz, A.~Zhou, {The fate
  of hints: updated global analysis of three-flavor neutrino oscillations},
  JHEP 09 (2020) 178.
\newblock \href {http://arxiv.org/abs/2007.14792} {\path{arXiv:2007.14792}},
  \href {http://dx.doi.org/10.1007/JHEP09(2020)178}
  {\path{doi:10.1007/JHEP09(2020)178}}.

\bibitem{Deppisch:2017xhv}
F.~F. Deppisch, T.~E. Gonzalo, L.~Graf, {Surveying the SO(10) Model Landscape:
  The Left-Right Symmetric Case}, Phys. Rev. D 96~(5) (2017) 055003.
\newblock \href {http://arxiv.org/abs/1705.05416} {\path{arXiv:1705.05416}},
  \href {http://dx.doi.org/10.1103/PhysRevD.96.055003}
  {\path{doi:10.1103/PhysRevD.96.055003}}.

\bibitem{Rothstein:1990qx}
I.~Z. Rothstein, {Renormalization group analysis of the minimal left-right
  symmetric model}, Nucl. Phys. B 358 (1991) 181--194.
\newblock \href {http://dx.doi.org/10.1016/0550-3213(91)90536-7}
  {\path{doi:10.1016/0550-3213(91)90536-7}}.

\bibitem{Staub:2013tta}
F.~Staub, {SARAH 4 : A tool for (not only SUSY) model builders}, Comput. Phys.
  Commun. 185 (2014) 1773--1790.
\newblock \href {http://arxiv.org/abs/1309.7223} {\path{arXiv:1309.7223}},
  \href {http://dx.doi.org/10.1016/j.cpc.2014.02.018}
  {\path{doi:10.1016/j.cpc.2014.02.018}}.

\bibitem{Barbieri:1999ma}
R.~Barbieri, P.~Creminelli, A.~Strumia, N.~Tetradis, {Baryogenesis through
  leptogenesis}, Nucl. Phys. B 575 (2000) 61--77.
\newblock \href {http://arxiv.org/abs/hep-ph/9911315}
  {\path{arXiv:hep-ph/9911315}}, \href
  {http://dx.doi.org/10.1016/S0550-3213(00)00011-0}
  {\path{doi:10.1016/S0550-3213(00)00011-0}}.

\bibitem{Antusch:2003kp}
S.~Antusch, J.~Kersten, M.~Lindner, M.~Ratz, {Running neutrino masses, mixings
  and CP phases: Analytical results and phenomenological consequences}, Nucl.
  Phys. B 674 (2003) 401--433.
\newblock \href {http://arxiv.org/abs/hep-ph/0305273}
  {\path{arXiv:hep-ph/0305273}}, \href
  {http://dx.doi.org/10.1016/j.nuclphysb.2003.09.050}
  {\path{doi:10.1016/j.nuclphysb.2003.09.050}}.

\bibitem{Giudice:2003jh}
G.~Giudice, A.~Notari, M.~Raidal, A.~Riotto, A.~Strumia, {Towards a complete
  theory of thermal leptogenesis in the SM and MSSM}, Nucl. Phys. B 685 (2004)
  89--149.
\newblock \href {http://arxiv.org/abs/hep-ph/0310123}
  {\path{arXiv:hep-ph/0310123}}, \href
  {http://dx.doi.org/10.1016/j.nuclphysb.2004.02.019}
  {\path{doi:10.1016/j.nuclphysb.2004.02.019}}.

\bibitem{GERDA:2020xhi}
M.~Agostini, et~al., {Final Results of GERDA on the Search for Neutrinoless
  Double-$\beta$ Decay}, Phys. Rev. Lett. 125~(25) (2020) 252502.
\newblock \href {http://arxiv.org/abs/2009.06079} {\path{arXiv:2009.06079}},
  \href {http://dx.doi.org/10.1103/PhysRevLett.125.252502}
  {\path{doi:10.1103/PhysRevLett.125.252502}}.

\bibitem{KamLAND-Zen:2022tow}
S.~Abe, et~al., {First Search for the Majorana Nature of Neutrinos in the
  Inverted Mass Ordering Region with KamLAND-Zen}\href
  {http://arxiv.org/abs/2203.02139} {\path{arXiv:2203.02139}}.

\bibitem{Barabash:2018fun}
A.~S. Barabash, {Possibilities of future double beta decay experiments to
  investigate inverted and normal ordering region of neutrino mass}, Front. in
  Phys. 6 (2019) 160.
\newblock \href {http://arxiv.org/abs/1901.11342} {\path{arXiv:1901.11342}},
  \href {http://dx.doi.org/10.3389/fphy.2018.00160}
  {\path{doi:10.3389/fphy.2018.00160}}.

\bibitem{Capozzi:2017ipn}
F.~Capozzi, E.~Di~Valentino, E.~Lisi, A.~Marrone, A.~Melchiorri, A.~Palazzo,
  {Global constraints on absolute neutrino masses and their ordering}, Phys.
  Rev. D 95~(9) (2017) 096014, [Addendum: Phys.Rev.D 101, 116013 (2020)].
\newblock \href {http://arxiv.org/abs/2003.08511} {\path{arXiv:2003.08511}},
  \href {http://dx.doi.org/10.1103/PhysRevD.95.096014}
  {\path{doi:10.1103/PhysRevD.95.096014}}.

\bibitem{DeSalas:2018rby}
P.~F. De~Salas, S.~Gariazzo, O.~Mena, C.~A. Ternes, M.~T\'ortola, {Neutrino
  Mass Ordering from Oscillations and Beyond: 2018 Status and Future
  Prospects}, Front. Astron. Space Sci. 5 (2018) 36.
\newblock \href {http://arxiv.org/abs/1806.11051} {\path{arXiv:1806.11051}},
  \href {http://dx.doi.org/10.3389/fspas.2018.00036}
  {\path{doi:10.3389/fspas.2018.00036}}.

\bibitem{CUPID:2019imh}
W.~R. Armstrong, et~al., {CUPID pre-CDR}\href {http://arxiv.org/abs/1907.09376}
  {\path{arXiv:1907.09376}}.

\bibitem{Armengaud:2019loe}
E.~Armengaud, et~al., {The CUPID-Mo experiment for neutrinoless double-beta
  decay: performance and prospects}, Eur. Phys. J. C 80~(1) (2020) 44.
\newblock \href {http://arxiv.org/abs/1909.02994} {\path{arXiv:1909.02994}},
  \href {http://dx.doi.org/10.1140/epjc/s10052-019-7578-6}
  {\path{doi:10.1140/epjc/s10052-019-7578-6}}.

\bibitem{LEGEND:2017cdu}
N.~Abgrall, et~al., {The Large Enriched Germanium Experiment for Neutrinoless
  Double Beta Decay (LEGEND)}, AIP Conf. Proc. 1894~(1) (2017) 020027.
\newblock \href {http://arxiv.org/abs/1709.01980} {\path{arXiv:1709.01980}},
  \href {http://dx.doi.org/10.1063/1.5007652} {\path{doi:10.1063/1.5007652}}.

\bibitem{nEXO:2017nam}
J.~B. Albert, et~al., {Sensitivity and Discovery Potential of nEXO to
  Neutrinoless Double Beta Decay}, Phys. Rev. C 97~(6) (2018) 065503.
\newblock \href {http://arxiv.org/abs/1710.05075} {\path{arXiv:1710.05075}},
  \href {http://dx.doi.org/10.1103/PhysRevC.97.065503}
  {\path{doi:10.1103/PhysRevC.97.065503}}.

\bibitem{nEXO:2018ylp}
S.~A. Kharusi, et~al., {nEXO Pre-Conceptual Design Report}\href
  {http://arxiv.org/abs/1805.11142} {\path{arXiv:1805.11142}}.

\bibitem{Baak:2012kk}
M.~Baak, M.~Goebel, J.~Haller, A.~Hoecker, D.~Kennedy, R.~Kogler, K.~Moenig,
  M.~Schott, J.~Stelzer, {The Electroweak Fit of the Standard Model after the
  Discovery of a New Boson at the LHC}, Eur. Phys. J. C 72 (2012) 2205.
\newblock \href {http://arxiv.org/abs/1209.2716} {\path{arXiv:1209.2716}},
  \href {http://dx.doi.org/10.1140/epjc/s10052-012-2205-9}
  {\path{doi:10.1140/epjc/s10052-012-2205-9}}.

\bibitem{Carmi:2012in}
D.~Carmi, A.~Falkowski, E.~Kuflik, T.~Volansky, J.~Zupan, {Higgs After the
  Discovery: A Status Report}, JHEP 10 (2012) 196.
\newblock \href {http://arxiv.org/abs/1207.1718} {\path{arXiv:1207.1718}},
  \href {http://dx.doi.org/10.1007/JHEP10(2012)196}
  {\path{doi:10.1007/JHEP10(2012)196}}.

\bibitem{ATLAS:2012yve}
G.~Aad, et~al., {Observation of a new particle in the search for the Standard
  Model Higgs boson with the ATLAS detector at the LHC}, Phys. Lett. B 716
  (2012) 1--29.
\newblock \href {http://arxiv.org/abs/1207.7214} {\path{arXiv:1207.7214}},
  \href {http://dx.doi.org/10.1016/j.physletb.2012.08.020}
  {\path{doi:10.1016/j.physletb.2012.08.020}}.

\bibitem{ATLAS:2015yey}
G.~Aad, et~al., {Combined Measurement of the Higgs Boson Mass in $pp$
  Collisions at $\sqrt{s}=7$ and 8 TeV with the ATLAS and CMS Experiments},
  Phys. Rev. Lett. 114 (2015) 191803.
\newblock \href {http://arxiv.org/abs/1503.07589} {\path{arXiv:1503.07589}},
  \href {http://dx.doi.org/10.1103/PhysRevLett.114.191803}
  {\path{doi:10.1103/PhysRevLett.114.191803}}.

\bibitem{ATLAS:2016neq}
G.~Aad, et~al., {Measurements of the Higgs boson production and decay rates and
  constraints on its couplings from a combined ATLAS and CMS analysis of the
  LHC pp collision data at $ \sqrt{s}=7 $ and 8 TeV}, JHEP 08 (2016) 045.
\newblock \href {http://arxiv.org/abs/1606.02266} {\path{arXiv:1606.02266}},
  \href {http://dx.doi.org/10.1007/JHEP08(2016)045}
  {\path{doi:10.1007/JHEP08(2016)045}}.

\bibitem{CMS:2012qbp}
S.~Chatrchyan, et~al., {Observation of a New Boson at a Mass of 125 GeV with
  the CMS Experiment at the LHC}, Phys. Lett. B 716 (2012) 30--61.
\newblock \href {http://arxiv.org/abs/1207.7235} {\path{arXiv:1207.7235}},
  \href {http://dx.doi.org/10.1016/j.physletb.2012.08.021}
  {\path{doi:10.1016/j.physletb.2012.08.021}}.

\bibitem{CMS:2014suk}
S.~Chatrchyan, et~al., {Evidence for the direct decay of the 125 GeV Higgs
  boson to fermions}, Nature Phys. 10 (2014) 557--560.
\newblock \href {http://arxiv.org/abs/1401.6527} {\path{arXiv:1401.6527}},
  \href {http://dx.doi.org/10.1038/nphys3005} {\path{doi:10.1038/nphys3005}}.

\bibitem{Brinckmann:2018owf}
T.~Brinckmann, D.~C. Hooper, M.~Archidiacono, J.~Lesgourgues, T.~Sprenger, {The
  promising future of a robust cosmological neutrino mass measurement}, JCAP 01
  (2019) 059.
\newblock \href {http://arxiv.org/abs/1808.05955} {\path{arXiv:1808.05955}},
  \href {http://dx.doi.org/10.1088/1475-7516/2019/01/059}
  {\path{doi:10.1088/1475-7516/2019/01/059}}.

\bibitem{Mohapatra:1979nn}
R.~N. Mohapatra, B.~Sakita, {SO(2n) Grand Unification in an SU(N) Basis}, Phys.
  Rev. D 21 (1980) 1062.
\newblock \href {http://dx.doi.org/10.1103/PhysRevD.21.1062}
  {\path{doi:10.1103/PhysRevD.21.1062}}.

\bibitem{Babu:1992ia}
K.~S. Babu, R.~N. Mohapatra, {Predictive neutrino spectrum in minimal SO(10)
  grand unification}, Phys. Rev. Lett. 70 (1993) 2845--2848.
\newblock \href {http://arxiv.org/abs/hep-ph/9209215}
  {\path{arXiv:hep-ph/9209215}}, \href
  {http://dx.doi.org/10.1103/PhysRevLett.70.2845}
  {\path{doi:10.1103/PhysRevLett.70.2845}}.

\bibitem{Anderson:1993fe}
G.~Anderson, S.~Raby, S.~Dimopoulos, L.~J. Hall, G.~D. Starkman, {A Systematic
  SO(10) operator analysis for fermion masses}, Phys. Rev. D 49 (1994)
  3660--3690.
\newblock \href {http://arxiv.org/abs/hep-ph/9308333}
  {\path{arXiv:hep-ph/9308333}}, \href
  {http://dx.doi.org/10.1103/PhysRevD.49.3660}
  {\path{doi:10.1103/PhysRevD.49.3660}}.

\end{thebibliography}

\end{document}